\documentclass[12pt]{article}
\pdfoutput=1
\usepackage{epsfig,amsfonts,amsthm}
\usepackage{amsmath,amssymb}
\usepackage[normalem]{ulem}
\usepackage{multirow}
\usepackage{hyperref}
\usepackage{array}
\newcommand{\be}{\begin{equation}}
\newcommand{\ee}{\end{equation}}
\newcommand{\bea}{\begin{eqnarray}}
\newcommand{\eea}{\end{eqnarray}}

\def\lsim{\mathrel{\rlap{\lower4pt\hbox{\hskip1pt$\sim$}}
    \raise1pt\hbox{$<$}}}        
\def\gsim{\mathrel{\rlap{\lower4pt\hbox{\hskip1pt$\sim$}}
    \raise1pt\hbox{$>$}}}        
    
\usepackage{amsmath}
\usepackage{amsfonts}
\usepackage{amssymb}
\usepackage{subfig}
\usepackage{wrapfig}
\usepackage{graphicx}

\def\beq{\begin{equation}}
\def\eeq{\end{equation}}
\def\bea{\begin{eqnarray}}
\def\eea{\end{eqnarray}}

\def\<{\left\langle}
\def\>{\right\rangle}

\newcommand{\bt}{\begin{tabular}}
\newcommand{\et}{\end{tabular}}

\usepackage{tabu}

\usepackage{slashed}
\usepackage{amssymb} 
\usepackage{float}

\usepackage[justification=centering]{caption}

\usepackage{tikz}
\usetikzlibrary{decorations.pathmorphing,decorations.markings}

\usepackage{graphicx}

\tikzset{
photon/.style={decorate, decoration={snake,amplitude=2pt, segment length=5pt}, draw=black},
particle/.style={draw=black, postaction={decorate}, decoration={markings,mark=at position .5 with {\arrow[draw=black]{>}}}},
antiparticle/.style={draw=black, postaction={decorate}, decoration={markings,mark=at position .5 with {\arrow[draw=black]{>}}}},
gluon/.style={decorate, draw=black, decoration={coil,amplitude=4pt, segment length=5pt}}
goldstone/.style={draw=green,postaction={decorate},decoration={markings,mark=at position .5 with {\arrow[draw=blue]{>}}}}
}

\definecolor{grey}{cmyk}{0,0,0,0.75}
\definecolor{tangerine}{cmyk}{0,0.5,1,0}
\definecolor{darkgreen}{cmyk}{1,0,1,0.23}
\definecolor{Red}{rgb}{1,0,0}
\definecolor{Blue}{rgb}{0,0,1}
\definecolor{Green}{rgb}{0,1,0}

\usepackage[
top    = 3.5cm,
bottom = 3.5cm,
left   = 3.5cm,
right  = 3.5cm]{geometry}

\allowdisplaybreaks[2]
\begin{document}
\bibliographystyle{OurBibTeX}

\title{\hfill ~\\[-30mm]
                  \textbf{Suppression of scalar mediated FCNCs in a  $SU(3)_c\times SU(3)_L\times U(1)_X$-model 
                }        }
\author{\\[-5mm]
Katri Huitu\footnote{E-mail: {\tt Katri.Huitu@helsinki.fi}},\ 
Niko Koivunen\footnote{E-mail: {\tt Niko.Koivunen@helsinki.fi}}\  
 \\
  \emph{\small Department of Physics and Helsinki Institute of Physics,}\\
  \emph{\small Gustaf H{\"a}llstr{\"o}min katu 2, FIN-00014 University of Helsinki, Finland}\\[4mm]}
\maketitle

\vspace*{-0.250truecm}
\begin{abstract}
\noindent
{
The models based on $SU(3)_C\times SU(3)_L\times U(1)_X$ gauge symmetry (331-models) have been advocated to explain the number of fermion  families. These models place one quark family to a different representation than the other two. The traditional 331-models are  plagued by scalar mediated quark  flavour changing neutral currents (FCNC) at tree-level. So far there has been no concrete mechanisms to suppress these  FCNCs in 331-models.  Recently it has been shown that the Froggatt-Nielsen mechanism can be incorporated into the 331-setting in an economical fashion (FN331-model). The FN331-model  explains both the number of fermion families in nature and their mass hierarchy simultaneously. In this work we study the Higgs mediated quark FCNCs in FN331-model. The flavour violating couplings of quarks are suppressed by the ratio of the  $SU(2)_L \times U(1)_Y$ and  $SU(3)_L\times U(1)_X$ breaking scales. We find that the $SU(3)_L\times U(1)_X$-breaking scale can be as low as 5 TeV  in order to pass the flavour bounds.
} 
\end{abstract}

\thispagestyle{empty}
\vfill
\newpage
\setcounter{page}{1}

\section{Introduction}
The Standard Model of particle physics (SM) \cite{Glashow:1961tr}-\cite{salam}  has been enormously successful in explaining experimental results. It however still leaves many important questions unaswered. One of these questions is the \emph{flavour problem}: the  SM leaves the number of fermion families and their mass hierarchy unexplained.  

Models based on $SU(3)_C\times SU(3)_L\times U(1)_X$ gauge group have been  considered  as a framework for  family structure that explains the number of fermion families in nature. The models with $SU(3)_C\times SU(3)_L\times U(1)_X$ gauge symmetry are collectively called 331-models.
The cancellation of gauge anomalies in 331-models differs substantially from the SM. 
In the traditional 331-models \cite{Singer:1980sw}-\cite{Nguyen:1998ui} the gauge anomalies cancel when the number of fermion triplets is equal to the number of antitriplets. This is only possible if the  number of fermion families is three. 

Eventhough the traditional 331-models explain the number of fermion families, they leave the fermion mass hierarchy unexplained. Recently this was remedied in \cite{Huitu:2017ukq}, where it was observed that the   Froggatt-Nielsen mechanism \cite{Froggatt:1978nt} can be economically incorporated into the 331-model  (FN331).  The Froggatt-Nielsen mechanism is one of the few known methods to explain the fermion mass hierarchy. The FN331-models thus simultaneously explain the number of fermion families and their mass hierarchy.

 The original  Froggatt-Nielsen mechanism extends the  Standard Model by additional flavour symmetry, under which the SM fermions are charged. In the simplest case the FN symmetry is a simple $U(1)$ or $Z_N$.  FN mechanism also introduces a complex scalar field, the \emph{flavon}. 
The flavour symmetry forbids the SM Yukawa couplings, but allows for the  following effective operator:
\begin{equation}
\mathcal{L}\supset c^{f}_{ij}\left(\frac{\phi}{\Lambda}\right)^{n_{ij}}\bar{f}_{L,i}H f_{R,j}+h.c.,
\end{equation} 
where $\phi$ is the flavon, $\Lambda$ is the scale of new physics,  $H$ the SM Higgs doublet and $f_i$ any SM fermion of flavour $i$. The power $n_{ij}$ is determined by the FN charge conservation. As the flavon acquires a VEV the  SM Yukawa couplings are generated as effective operators. 
The FN331-model uses the existing scalar content of a 331-model to emulate the flavon in the Froggatt-Nielsen mechanism, in contrast to the FN mechanism in SM where the scalar sector has to be extended. This is interesting since the 331-models can  explain the number of fermion families. 

The 331-models contain one additional diagonal generator compared to the SM. There is therefore freedom in the way how the electric charge is embedded into the  $SU(3)_C\times SU(3)_L\times U(1)_X$. The electric charge in 331-models can be written in a general form as:
\begin{equation}
Q=T_3+\beta T_8+X,
\end{equation}
where $T_3$ and $T_8$ are the diagonal $SU(3)_L$ generators and $X$ is the $U(1)_X$ charge. The choice of parameter $\beta$ defines the type of the 331-model. Models with $\beta=\pm 1/\sqrt{3}$ are well studied in the literature \cite{Singer:1980sw}-\cite{Dong:2013ioa}. These models do not contain particles with exotic electric charges and the scalar sector is relatively simple.  The models with $\beta=\pm 1/\sqrt{3}$ require minimally  three scalar triplets, $\eta=(\eta_1^+,\eta^0,\eta_2^+)^T$, $\rho=(\rho_1^0,\rho^-,\rho_2^0)^T$ and $\chi=(\chi_1^0,\chi^-,\chi_2^0)^T$,   to break the gauge symmetries and to generate tree-level masses for all the gauge bosons and fermions (except neutrinos).  An important property of this minimal  scalar sector  is the fact that it inevitably contains two triplets in the same representation, $\rho$ and $\chi$ in this case. The FN331-model is based on choice $\beta=\pm 1/\sqrt{3}$ due to this  special property of its scalar sector: combination $\rho^\dagger \chi$ can act as an effective flavon \cite{Huitu:2017ukq}.   The combination $\rho^\dagger \chi$ is gauge invariant and can carry a non-zero  $U(1)$-charge. The most general electric charge conserving vacuum is  $\langle\eta\rangle=(0,v_\eta,0)$, $\langle\rho\rangle=(v_{\rho_1},0,v_{\rho_2})$ and $\langle\chi\rangle=(0,0,v_\chi)$. Therefore, the combination  $\rho^\dagger \chi$ has a non-zero VEV.  Similar mechanism has been studied in the 2HDM models \cite{Bauer:2015fxa},\cite{Bauer:2015kzy}, where the effective flavon is comprised of two Higgs doublets.

The models with only two scalar triplets have been studied in $\beta=\pm 1/\sqrt{3}$ \cite{Ponce:2002sg}, \cite{Dong:2006mg}. The models with this reduced scalar sector don't generate all the fermion masses at tree-level and  rely on radiative corrections  to generate  the masses of the lightest fermions.

Also the models with $\beta=\pm \sqrt{3}$ have been studied in the literature \cite{Pisano:1991ee}-\cite{Nguyen:1998ui}. 
The scalar sector in those models is  quite complicated, since the generation of  masses to all of the charged fermions at tree-level  requires three scalar triplets and one scalar sextet.
 These models also contain  particles with exotic electric charges, such as doubly charged bosons and quarks with electric charges $\pm 5/3$ and $\pm 4/3$.  The three scalar triplets are all in different representations and one cannot construct a gauge invariant combination that could carry a non-zero charge. Therefore, the Froggatt-Nielsen mechanism cannot  economically be incorporated into the models with $\beta=\pm\sqrt{3}$.

The anomaly cancellation produces  the most appealing property of the 331-models: the explanation for the  number of families. But it also causes the most unpleasant feature to the 331-models: one of the quark generations has to be placed into different representation than the two other generations. The quarks will as a result couple to multiple scalar triplets. This inevitably leads to tree-level  scalar mediated flavour changing neutral currents  for quarks \cite{Glashow:1976nt},\cite{Paschos:1976ay}. This is a general feature that plagues the 331-models. The traditional 331-models   \cite{Singer:1980sw}-\cite{Nguyen:1998ui} do not offer any natural suppression mechanism to the tree-level scalar mediated quark FCNCs. In this work we study the suppression of tree-level scalar mediated FCNCs of quarks in FN331-model.  The FN331-model employs the FN mechanism for the generation of the Yukawa couplings and the form of the fermion diagonalization matrices is therefore known.  This allows us to study the flavour violating  Yukawa couplings of quarks  in great detail.   We find  the FCNCs to be naturally suppressed in FN331-model. 

The paper is structured as follows. In Section \ref{particle content} we present the full particle content of the FN331-model and the scalar sector is examined more closely in Section \ref{scalar masses}. The details of the  Froggatt-Nielsen mechanism are presented in Section \ref{Yukawa sector}, after which we take a closer look on the fermion masses and the quark mass matrices in particular in Section \ref{Yukawa couplings and fermion masses}. The Higgs mediated FCNCs of quarks are examined in Section  \ref{scalar FCNC} after which we discuss the  constraints imposed by the neutral meson mixing in Section \ref{Higgs mediated neutral meson mixing}. Finally in Section \ref{numerical examples} we present three numerical examples with different quark mass textures to illustrate the suppression of FCNCs.

\section{Particle content}\label{particle content}
In FN331-model  the gauge group of the Standard Model  is extended to $SU(3)_c\times SU(3)_L\times U(1)_X$. We define the electric charge as\footnote{The choice $\beta=+\frac{1}{\sqrt{3}}$ would result in essentially the same model.}:

\begin{equation}
Q=T_3+\beta T_8+X=T_3-\frac{1}{\sqrt{3}}T_8+X,
\end{equation}
where the $T_3$ and $T_8$ are the diagonal $SU(3)_L$ generators. 

We also introduce a  global $U(1)_{FN}$-symmetry, under which fermions and some of the scalars are charged. The $U(1)_{FN}$-symmetry will be spontaneously broken by the $SU(3)_L$-vacuum. This will generate the fermion mass hierarchy through the Froggatt-Nielsen mechanism.

\subsection{Fermion representations}\label{fermion representations}
 Let us now write down the fermion representations. The SM fermions have to be assigned into specific representations in order to cancel the gauge anomalies, as already stated in the introduction.  The left-handed leptons are assigned to  $SU(3)_{L}$ -triplets and the right-handed charged leptons are assigned to $SU(3)_{L}$-singlets:

\begin{eqnarray}
&&L_{L,i}=\left(\begin{array}{c}
\nu_{i}\\
e_{i}\\
\nu'_{i}
\end{array}
\right)_{L}\sim (1,3,-\frac{1}{3}), \\
&&  e_{R,i}\sim (1,1,-1)\quad i=1,2,3. 
\end{eqnarray}
 The numbers in the parantheses label the transformation properties under the gauge group  $SU(3)_{c}\times SU(3)_{L}\times U(1)_{X}$. The $\nu'_{L,i}$ 
 are new leptons with electric charge $0$. We have not introduced right-handed neutrino-like singlets. The details of the anomaly cancellation do not depend on their precence, as they are gauge singlets.  Models with $\beta=\pm1/\sqrt{3}$ without the neutrino-like singlets have one massless neutrino and mass degenerate Dirac-neutrinos at tree-level. Loop corrections are needed to break  the degeneracy and  lift the one mass from zero \cite{Valle:1983dk}. In the models where the right-handed singlets are present produce the non-zero neutrino masses at tree-level for all the neutrinos.  

The cancellation of anomalies requires the number of fermion triplets to be the same as antitriplets. This is achieved by assigning two quark families to $SU(3)_{L}$ antitriplets and one family to a triplet. We choose to assign the  first quark generation into triplet and the second and the third  into an antitriplet:

\begin{eqnarray}
&&Q_{L,1}=\left(\begin{array}{c}
u_{1}\\
d_{1}\\
U
\end{array}
\right)_{L}\sim (3,3,\frac{1}{3}),\\
&&Q_{L,2}=\left(\begin{array}{c}
d_{2}\\
-u_{2}\\
D_{1}
\end{array}
\right)_{L},\quad 
Q_{L,3}=\left(\begin{array}{c}
d_{3}\\
-u_{3}\\
D_{2}
\end{array}
\right)_{L}\sim (3,3^{\ast},0),\\
&&  u_{R,i}\sim (3,1,\frac{2}{3}), \quad  U_{R}\sim (3,1,\frac{2}{3}),\\
&&  d_{R,i}\sim (3,1,-\frac{1}{3}), \quad  D_{R,1}\sim (3,1,-\frac{1}{3}), \quad  D_{R,2}\sim (3,1,-\frac{1}{3}),\quad i=1,2,3.
\end{eqnarray}
 
We have introduced new quarks $D_{1}$ and  $D_{2}$ with electric charge $-1/3$ and $U$ with electric  charge $2/3$. When we take into account the colour, there are six fermion triplets and six antriplets, ensuring the cancellation of pure $SU(3)_L$-anomaly. All the gauge anomalies will cancel with this particle content.

The fermions are also charged under global Froggatt-Nielsen $U(1)_{FN}$ symmetry. This will be discussed in detail in Section \ref{Yukawa sector}.

\subsection{Scalar sector}
As already stated in the introduction the key feature of our model is the fact that the  331-models with $\beta=\pm\frac{1}{\sqrt{3}}$ contain only two types of scalar triplets with neutral entries: $X=2/3$ and $X=-1/3$. One must include at least  two  triplets with $X=-1/3$  and one  triplet with $X=2/3$ in order to generate the masses for all the charged fermions at the tree-level (as will be discussed later). We choose to have this minimal scalar sector:

\begin{eqnarray}
&&\eta=\left(\begin{array}{c}
\eta^{+}\\
\eta^{0}\\
{\eta'}^{+}
\end{array}
\right)\sim (1,3,\frac{2}{3}),\quad  
\rho=\left(\begin{array}{c}
\rho^{0}\\
\rho^{-}\\
{\rho'}^{0}
\end{array}
\right),\quad \chi=\left(\begin{array}{c}
\chi^{0}\\
\chi^{-}\\
{\chi'}^{0}
\end{array}
\right)\sim (1,3,-\frac{1}{3}).\label{scalar triplets}
\end{eqnarray}
 
All  the neutral fields can in general develop non-zero VEV.  The minima are, however, related to each other  by $SU(3)_L$ rotation. We can therefore choose to rotate  one of the $X=-1/3$ triplet VEVs so that the upper component VEV goes to zero. This rotation will leave the rest of the VEVs general. Thus the most general vacuum structure is:  

\begin{eqnarray}
&&\langle\eta\rangle=\frac{1}{\sqrt{2}}\left(\begin{array}{c}
0\\
v'\\
0
\end{array}
\right), \quad
\langle\rho\rangle=\frac{1}{\sqrt{2}}\left(\begin{array}{c}
v_1\\
0\\
v_2
\end{array}
\right),
\quad
\langle\chi\rangle=\frac{1}{\sqrt{2}}\left(\begin{array}{c}
0\\
0\\
u
\end{array}
\right).\label{vacuum}
\end{eqnarray}  
The VEVs $v_2$ and $u$ break the $SU(3)_L\times U(1)_X\to SU(2)_L\times U(1)_Y$.  
The VEVs $v'$ and $v_1$ break the $SU(2)_L\times U(1)_Y\to U(1)_{\rm em}$  and we take them to be of the order of electroweak scale.  We assume $v_2,u >> v',v_1$.

The scalar triplets in Eq.  (\ref{scalar triplets}) are charged under the global symmetry $U(1)_{FN}$. The charge assignments are presented in the Table \ref{FN charges of scalars}.
 The scalar potential is greatly simplified due to inclusion of global $U(1)_{FN}$-symmetry.

\begin{table}[h!]
\begin{center}
\begin{tabular}{|c|c|c|c|}
\hline
Particle   &  $\eta$ & $\rho$ & $\chi$ \\
\hline 
FN-charge & $-1$ & $1$ & $0$ \\
\hline
\end{tabular}
\end{center}
\caption{The FN $U(1)$ charges of the scalars.}
\label{FN charges of scalars}
\end{table}

 The most general $U(1)_{FN}$-symmetric scalar potential is\footnote{Note that the only term  the potential in Eq.  (\ref{FN symmetric potential}) that could be forbidden by $U(1)_{FN}$-charge assignment is the trilinear term in the last line. The trilinear term is needed to generate mass to all scalars, except the Goldstone bosons.  
 }:

\begin{eqnarray}
&& V_{\textrm{FN}}=\mu^2_1 \eta^{\dagger}\eta+\mu_2^2 \rho^{\dagger}\rho+\mu_3^{2}\chi^{\dagger}\chi
+\lambda_1 (\eta^{\dagger}\eta)^2+\lambda_2 (\rho^{\dagger}\rho)^2+\lambda_{3}(\chi^{\dagger}\chi)^2\label{FN symmetric potential}\\
&& +\lambda_{12} (\eta^{\dagger}\eta)(\rho^{\dagger}\rho)+\lambda_{13} (\eta^{\dagger}\eta)(\chi^{\dagger}\chi)+\lambda_{23}(\rho^{\dagger}\rho)(\chi^{\dagger}\chi)\nonumber\\
&& +\widetilde{\lambda}_{12} (\eta^{\dagger}\rho)(\rho^{\dagger}\eta)+\widetilde{\lambda}_{13} (\eta^{\dagger}\chi)(\chi^{\dagger}\eta)+\widetilde{\lambda}_{23}(\rho^{\dagger}\chi)(\chi^{\dagger}\rho)\nonumber\\
&&+\sqrt{2}f(\epsilon_{ijk}\eta^{i}\rho^{j}\chi^{k}+h.c.).\nonumber
\end{eqnarray}

The scalar field VEVs break the global $U(1)_{\rm FN}$ symmetry spontaneously, and 
 this leaves one Golstone boson to the physical spectrum. In order to give it  a mass we add the following soft FN-breaking term to the potential:
\begin{equation}
V_{\textrm{soft}}=b(\rho^{\dagger}\chi)+h.c.
 \end{equation}
The full scalar potential is:
\be\label{full scalar potential}
V=V_{\rm FN}+V_{\rm soft}.
\ee

The minimization conditions for the potential $V$ are:
\begin{eqnarray}
\mu_1^2 & = & -\lambda_1 {v'}^2-\frac{1}{2}\lambda_{12}(v_1^2+v_2^2)-\frac{1}{2}\lambda_{13}u^2+f\frac{v_1 u}{v'},\nonumber\\
\mu_2^2 & = & -\lambda_2 ({v_1}^2+v_2^2)-\frac{1}{2}\lambda_{12}{v'}^2-\frac{1}{2}\lambda_{23}u^2+f\frac{v'  u}{v_1},\nonumber\\
\mu_3^2 & = & -b\frac{v_2}{u}-\lambda_3 {u}^2-\frac{1}{2}\lambda_{13}{v'}^2-\frac{1}{2}\lambda_{23}(v_1^2+v_2^2)-\frac{1}{2}\widetilde{\lambda}_{23}v_2^2+f\frac{v'  v_1}{u},\nonumber\\
\frac{1}{2}\widetilde{\lambda}_{23}u^2&=&-b\frac{u}{v_2}-f \frac{v' u}{v_1}.\nonumber
\end{eqnarray}
The complex phases in the couplings $f$ and $b$ can be absorbed into the fields and therefore all the parameters in the potential $V$ are real.

\subsection{Gauge sector}

The covariant derivative for triplet is:
\begin{eqnarray}
&&D_{\mu}=\partial_{\mu}-ig_3 \sum_{a=1}^8 T_{a}W_{a\mu}-ig_{x}XB_{\mu},\nonumber
\end{eqnarray}
where  $g_3$ and $g_x$ are the $SU(3)_L$ and $U(1)_X$ gauge couplings, respectively. The $T_a=\lambda_a/2$ are the $SU(3)_L$ generators, where $\lambda_a$ are the Gell-Mann matrices. 

The $SU(3)_L$ gauge bosons are: 
\begin{eqnarray}
 &&\sum_{a=1}^8 T_a W_{a\mu}=\frac{1}{\sqrt{2}}\left(\begin{array}{ccc}
\frac{1}{\sqrt{2}}W_{3\mu}+\frac{1}{\sqrt{6}}W_{8\mu} &
 {W'}_{\mu}^{+} &  
{X'}^{0}_{\mu}\\
 {W'}^{-}_{\mu} &
-\frac{1}{\sqrt{2}}W_{3\mu}+\frac{1}{\sqrt{6}}W_{8\mu}& 
 {V'}^{-}_{\mu}\\
 X'^{0\ast}_{\mu} & 
 {V'}^{+}_{\mu}&
-\frac{2}{\sqrt{6}}W_{8\mu}
\end{array}\right),\nonumber
\end{eqnarray} 
where we have denoted;
\begin{eqnarray}
{W'}^{\pm}_\mu=\frac{1}{\sqrt{2}}(W_{2\mu}\mp iW_{3\mu}),\nonumber\\
{V'}^{\mp}_\mu=\frac{1}{\sqrt{2}}(W_{6\mu}\mp iW_{7\mu}),\nonumber\\
{X'}^{0}_\mu=\frac{1}{\sqrt{2}}(W_{4\mu}-iW_{5\mu}).\nonumber
\end{eqnarray}

The off-diagonal neutral  gauge bosons will mix with the diagonal ones, due to two non-zero VEVs in scalar triplet $\rho$.   
The fields  $W_{3\mu}$, $W_{8\mu}$, $B_\mu$ and $W_{4\mu}$ will form neutral mass eigenstates: photon, $Z$-boson and new heavy gauge bosons $Z'$ and $\mbox{\footnotesize $\widetilde{W}_{4\mu}$}$. The field $W_{5\mu}$ does not mix with the other neutral gauge bosons and is a mass eigenstate, with same mass as $\mbox{\footnotesize $\widetilde{W}_{4\mu}$}$. These fields are identified as a \emph{physical neutral non-hermitian gauge boson} $X^0_\mu\equiv\frac{1}{\sqrt{2}}(\widetilde{W}_{4\mu}-iW_{5\mu})$ \cite{Ponce:2002sg},\cite{Dong:2006mg}. The off-diagonal gauge bosons ${W'}^{\pm}_\mu$ and ${V'}^{\pm}_\mu$ will mix and  form the SM gauge bosons $W^{\pm}_\mu$ and the heavy new gauge bosons $V^{\pm}_\mu$. The masses of the new gauge bosons are proportional to the $SU(3)_L\times U(1)_X$-breaking VEVs and the new gauge bosons are presumably very heavy. The expressions of the new gauge boson masses are given in the Appendix \ref{gauge boson masses appendix}, where further details of the gauge sector are provided.

The masses of the SM gauge bosons are given as:
\begin{eqnarray}
&&m_W^2=  \frac{g_3^2}{4}({v'}^2+\frac{v_1^2 u^2}{v_2^2+u^2})+\mathcal{O}\left(\frac{v^2_{\textrm{light}}}{v_{\textrm{heavy}}^2}\right),\nonumber\\
&& m_Z^2 = \frac{g_3^2}{4}\left(\frac{3g_3^2+4g_x^2}{3g_3^2+g_x^2}\right)\left({v'}^2+\frac{v_1^2  u^2}{v_2^2+u^2}\right)+\mathcal{O}\left(\frac{v^2_{\textrm{light}}}{v_{\textrm{heavy}}^2}\right),\nonumber
\end{eqnarray}
where $v_{\textrm{light}}=v',v_1$,  $v_{\textrm{heavy}}=u, v_2$ and 
\begin{equation}
\cos^2 \theta_W=\frac{3g_3^2+g_x^2}{3g_3^2+4g_x^2}
\end{equation}
defines the Weinberg angle.
At low energies we identify the $g_3$  with the SM $SU(2)_L$ gauge coupling. The SM Higgs VEV is related to the triplet VEVs through the relation
\begin{equation}\label{relation among vevs}
{v'}^2+\frac{v_1^2 u^2}{v_2^2+u^2}+\mathcal{O}\left(\frac{v^2_{\textrm{light}}}{v^2_{\textrm{heavy}}}\right)=(v_{sm})^2,
\end{equation}
where $v_{sm}=246 \textrm{ GeV}$.

\section{Scalar masses}\label{scalar masses}
The neutral scalars are divided to real and imaginary parts as:
\begin{eqnarray}
&&\eta^{0}=\frac{1}{\sqrt{2}}(v'+h_1+i \xi_1),\nonumber\\
&&\rho^{0}=\frac{1}{\sqrt{2}}(v_1+h_2+i \xi_2), \quad {\rho'}^{0}=\frac{1}{\sqrt{2}}(v_2+h_4+i \xi_4),\nonumber\\
&&\chi^{0}=\frac{1}{\sqrt{2}}(h_3+i \xi_3), \quad {\chi'}^{0}=\frac{1}{\sqrt{2}}(u+h_5+i \xi_5).\nonumber
\end{eqnarray}

All the parameters of the scalar potential are real and  therefore the real and imaginary parts of the scalars do not mix. The CP-even and the CP-odd scalars form their own  $5\times 5$ mass matrices and the charged scalars form $4\times 4$ mass matrix.

\subsection{CP-even scalars}
The CP-even scalar  mass term is,
\be\nonumber
 \mathcal{L}\supset \frac{1}{2}H^T M_{cp-even}^2 H,
\ee
where  $H^T=(h_1,h_2,h_3,h_4,h_5)$ and, 
\begin{eqnarray}
\resizebox{1.0 \textwidth}{!} 
{
$ 
M_{cp-even}^2=
\left(\begin{array}{ccccc}
2\lambda_1{v'}^2 +f\frac{v_1 u}{v'} & \lambda_{12}v' v_1 -fu & fv_2 & \lambda_{12} v' v_2 &  \lambda_{13} v' u-f v_1\\
\lambda_{12}v' v_1 -fu   &2\lambda_2{v_1}^2 +f\frac{v'  u}{v_1} &\frac{1}{2}\widetilde{\lambda}_{23} v_2 u +b & 2\lambda_2 v_1 v_2 & \lambda_{23} v_1 u -fv'\\
fv_2 & \frac{1}{2}\widetilde{\lambda}_{23} v_2 u +b & -\frac{1}{2}\widetilde{\lambda}_{23} v_{2}^2 -b\frac{(v_1^2+v_2^2)}{v_2 u} & -b\frac{v_1}{v_2} & \frac{1}{2}\widetilde{\lambda}_{23}v_1 v_2\\
 \lambda_{12} v' v_2&2\lambda_2 v_1 v_2  &-b\frac{v_1}{v_2} &2\lambda_2 v_2^2 -b\frac{u}{v_2} &(\lambda_{23} +\widetilde{\lambda}_{23})v_2 u +b\\
\lambda_{13} v' u-f v_1&\lambda_{23} v_1 u -fv' &  \frac{1}{2}\widetilde{\lambda}_{23}v_1 v_2 &(\lambda_{23} +\widetilde{\lambda}_{23})v_2 u +b&2\lambda_3{u}^2 +f\frac{v'  v_1}{u}-b\frac{v_2}{u} 
\end{array}\right).\label{cp-even mass matrix}
$
}
\end{eqnarray}
The matrix is diagonalized as:
\begin{equation}\label{cp-even eigenstates}
U^{H\dagger}M_{CP-even}^2 U^{H}=M_{\textrm{H}}^2=\textrm{diag}(0,m_h^2,m_{H_1}^2,m_{H_2}^2,m_{H_3}^2).
\end{equation}
This matrix has one zero eigenvalue, corresponding to the Goldstone boson that gives mass to the  neutral non-Hermitian gauge boson $X^0_\mu$ and four non-zero eigenvalues corresponding to four physical CP-even scalars $h$, $H_1$, $H_2$ and $H_3$. One of the non-zero eigenvalues is $\mathcal{O}(v_{\rm sm}^2)$ and is identified with the 125 GeV Higgs boson $h$ of the SM and therefore $m_h^2=(125 \textrm{ GeV})^2$.  
The three of the eigenvalues are $\mathcal{O}(v_{\rm heavy}^2)$ and therefore very heavy. The heavy eigenvalues to the leading order are:
\begin{eqnarray}
m_{H_1}^2&=&\frac{1}{2}\left(-\widetilde{\lambda}_{23}-\frac{2b}{u v_2}\right)\left[\frac{v_1^2}{{v'}^2}u^2+u^2+v_2^2\right]+\mathcal{O}\left(\frac{v^2_{\textrm{light}}}{v^2_{\textrm{heavy}}}\right),\nonumber\\
m_{H_2}^2&=&A-\sqrt{A^2-B}+\mathcal{O}\left(\frac{v^2_{\textrm{light}}}{v^2_{\textrm{heavy}}}\right),\nonumber\\
m_{H_3}^2&=&A+\sqrt{A^2-B}+\mathcal{O}\left(\frac{v^2_{\textrm{light}}}{v^2_{\textrm{heavy}}}\right),\nonumber
\end{eqnarray}
where,
\begin{eqnarray}
A&=&\lambda_2  v_2^2+\lambda_3 u^2 -\frac{b}{2}\frac{ u^2+ v_2^2}{v_2 u},\nonumber\\
B&=&\left[4\lambda_2 \lambda_3 -(\widetilde{\lambda}_{23}+\lambda_{23})^2\right]u^2 v_2^2-2b(\widetilde{\lambda}_{23}+\lambda_{23})u v_2-\frac{2b}{v_2 u}(\lambda_2 v_2^4+\lambda_3 u^4).\nonumber
\end{eqnarray}

 The CP-even mass eigenstates are defined as:
\begin{equation}\label{CP-even rotation}
\left(\begin{array}{c}
G^0_1\\
h\\
H_1\\
H_2\\
H_3
\end{array}\right)
=U^{H\dagger} \left(\begin{array}{c}
h_1\\
h_2\\
h_3\\
h_4\\
h_5
\end{array}\right),
\end{equation}
where $G^0_1$ is the Goldstone boson giving the mass to the $X_\mu^0$ gauge boson.

We are mainly interested of the interactions of the lightest neutral scalar, we assume to be the Higgs as this contributes to the FCNCs the most.  The eigenvector corresponding to Higgs\footnote{$M^2_{cp-even}\bar{X}_{h}=m_h^2\bar{X}_{h}$} at the leading order is:
\begin{equation}\label{higgs eigenvector approximation}
\bar{X}_{h}=
\left(
\begin{array}{c}
U^{H}_{12}\\
U^{H}_{22}\\
U^{H}_{32}\\
U^{H}_{42}\\
U^{H}_{52}\\
\end{array}\right)=
\left(
\begin{array}{c}
\frac{v'}{v_{sm}}+\mathcal{O}\left(\frac{v_{\textrm{light}}}{v_{\textrm{heavy}}}\right)\\
\frac{v_{sm}^2-{v'}^2}{v_{sm}v_1}+\mathcal{O}\left(\frac{v_{\textrm{light}}}{v_{\textrm{heavy}}}\right)\\
-\frac{v_2}{u}\frac{v_{sm}^2-{v'}^2}{v_{sm}v_1}+\mathcal{O}\left(\frac{v_{\textrm{light}}}{v_{\textrm{heavy}}}\right)\\
0+\mathcal{O}\left(\frac{v_{\textrm{light}}}{v_{\textrm{heavy}}}\right)\\
0+\mathcal{O}\left(\frac{v_{\textrm{light}}}{v_{\textrm{heavy}}}\right)\\
\end{array}\right).
\end{equation}
This explicit form of the eigenvector will be important when discussing the FCNCs is Section \ref{scalar FCNC}.

\subsection{CP-odd scalars}
The CP-odd scalar  mass term is,
\be\nonumber
 \mathcal{L}\supset \frac{1}{2}A^T M_{cp-odd}^2 A,
\ee
where  $A^T=(\xi_1,\xi_2,\xi_3,\xi_4,\xi_5)$ and, 
\begin{eqnarray}
M_{cp-odd}^2=
\left(\begin{array}{ccccc}
f\frac{v_1 u}{v'} & fu & -f v_2 & 0 &  fv_1\\
fu   &f\frac{v'  u}{v_1} & -\frac{f v_2 v'}{v_1} & 0 & fv'\\
-f v_2& -\frac{f v_2 v'}{v_1} &  -\frac{b v_1^2}{u v_2}+\frac{f v_2^2 v'}{u v_1} &  b\frac{v_1}{v_2}  & -\frac{b v_1+f v_2 v'}{u}\\
0 & 0  &b\frac{v_1}{v_2} &  -b \frac{u}{v_2} & b  \\
fv_1& fv' &  -\frac{b v_1+f v_2 v'}{u} & b &  f\frac{v'  v_1}{u}-b\frac{v_2}{u} 
\end{array}\right).\nonumber
\end{eqnarray}
This matrix has three zero eigenvalues corresponding to the Goldstone bosons that give masses to $Z$, $Z'$ and the neutral non-Hermitian gauge boson $X^0_\mu$. The two non-zero eigenvalues correspond to physical CP-odd scalars $A_1$ and $A_2$. 
One of the non-zero eigenvalues is proportional to $SU(3)_L\times U(1)_X$-breaking scale and the other one depends on the soft  Froggatt-Nielsen symmetry breaking term $b$.
The pseudo-scalar mass matrix is diagonalized as: 
\begin{equation}
U^{A\dagger}M_{CP-odd}^2 U^{A}=M_{\textrm{A}}^2=\textrm{diag}(m_{A_1}^2,m_{A_2}^2,0,0,0) ,
\end{equation}
where
\begin{eqnarray}
&&m_{A_1}^2=\frac{1}{2}\left(-\widetilde{\lambda}_{23}-\frac{2b}{u v_2}\right)\left[\frac{v_1^2}{{v'}^2}u^2+u^2+v_2^2+v_1^2\right],\nonumber\\
&&m_{A_2}^2= -\frac{b(u^2 +v_2^2+v_1^2)}{u v_2}.\nonumber
\end{eqnarray}
The mass of the CP-odd scalar $A_2$ is proportional to the explicit $U(1)_{FN}$-breaking term, and is, therefore, identified as a pseudo-Goldstone boson.

\subsection{Charged  scalars}
The charged scalar  mass term is,
\be\nonumber
 \mathcal{L}\supset C^T M_{\textrm{charged scalar}}^2 C,
\ee
where  $C^T=( {\eta'}^{+},\eta^{+},\rho^{+},\chi^{+})$ and,
\begin{eqnarray}
\resizebox{1.0 \textwidth}{!} 
{
$ 
M_{\rm charged}^2=
\left(\begin{array}{cccc}
 f\frac{v_1 u}{v'}+\frac{1}{2}\widetilde{\lambda}_{12}v_2^2 + \frac{1}{2}\widetilde{\lambda}_{13}u^2   & \frac{1}{2}\widetilde{\lambda}_{12}v_1 v_2 & \frac{1}{2}\widetilde{\lambda}_{12}v' v_2 & \frac{1}{2}\widetilde{\lambda}_{13} v'  u +f v_1 \\
 \frac{1}{2}\widetilde{\lambda}_{12}v_1 v_2&  f\frac{v_1 u}{v'}+\frac{1}{2}\widetilde{\lambda}_{12}v_1^2  &  \frac{1}{2}\widetilde{\lambda}_{12}v'  v_1 +fu& -fv_2\\
\frac{1}{2}\widetilde{\lambda}_{12}v' v_2  &  \frac{1}{2}\widetilde{\lambda}_{12}v'  v_1 +fu & f\frac{v'  u}{v_1}+\frac{1}{2}\widetilde{\lambda}_{12}{v'}^2 &   \frac{1}{2}\widetilde{\lambda}_{23} v_2  u+b \\
 \frac{1}{2}\widetilde{\lambda}_{13} v'  u +f v_1 & -fv_2 & \frac{1}{2}\widetilde{\lambda}_{23} v_2  u +b&   f\frac{v' v_1}{u}-\frac{1}{2}\widetilde{\lambda}_{23}v_2^2 + \frac{1}{2}\widetilde{\lambda}_{13}{v'}^2-b\frac{v_2}{u}
\end{array}\right).\nonumber
$
}
\end{eqnarray}
The matrix has two zero eigenvalues corresponding to the Goldstone bosons giving mass to the $W^{\pm}_\mu$ and $V^{\pm}_\mu$  gauge bosons. The two  non-zero eigenvalues correspond to physical charged scalars $H_1^+$ and  $H_2^+$. The charged scalar mass matrix is diagonalized as:
\begin{equation}
U^{C\dagger}M_{\rm charged}^2 U^{C}=M_{\textrm{C}}^2=\textrm{diag}(0,0,m_{H_1^+}^2,m_{H_2^+}^2) ,
\end{equation}
The masses of the physical charged scalars are:
\bea
&&m_{H_1}^2=\frac{1}{2}\left[\left(-\widetilde{\lambda}_{23}-\frac{2b}{uv_2}\right)\frac{u^2 v_1^2}{{v'}^2}+\left(\widetilde{\lambda}_{12}v_2^2+\widetilde{\lambda}_{13}u^2\right)\right]\\
&&m_{H_2}^2=\frac{1}{2}\left(-\widetilde{\lambda}_{23}-\frac{2b}{u v_2}\right)\left(\frac{u^2 v_1^2}{{v'}^2}+u^2+v_2^2\right).
\eea
Both the physical charged scalar masses are proportional  to $SU(3)_L\times U(1)_X$-breaking scale and are assumed to be heavy.

\section{Froggatt-Nielsen mechanism}\label{Yukawa sector}
For  the Yukawa sector of the model, we  employ the Froggatt-Niesen mechanism to generate the fermion mass hierarchy. First we review the original Froggatt-Nielsen framework in Section \ref{original FN}, then we formulate  the Froggatt-Nielsen mechanism in the 331-framework in the Section \ref{FN in 331}.

\subsection{Review of the original Froggatt-Nielsen framework}\label{original FN}

The  original Froggatt-Nielsen model extends the Standard Model with a flavour symmetry (FN symmetry), whose symmetry group in the simplest case is global or local $U(1)$ or  a discrete $Z_N$ symmetry. New fields are  introduced: heavy fermion messengers $\xi_a$, and complex scalar field $\phi$, called the flavon, which is a $SU(3)_C\times SU(2)_L\times U(1)_Y$ singlet. The SM fermions, the  SM Higgs and the new particles are charged under the FN symmetry as presented in the Table \ref{FN charges}.

\begin{table}[h!]
\begin{center}
\begin{tabular}{|c|c|c|c|c|c|}
\hline
Particle    & $\phi$ & $(\psi^f_{L,i})^{c}$ & $f_{R,i}$ & $H$ & $\xi_a$\\
\hline 
FN charge  & $q_{\phi}$ & $q_{\bar L, i}$ & $q_{R,i}$ & $q_{H}$ & $q_{\xi_a}$ \\
\hline
\end{tabular}
\end{center}
\caption{The FN charges of  fermions and the scalar fields.}
\label{FN charges}
\end{table}

 The FN  charge assignment is  such that the SM Yukawa couplings are forbidden (with the possible exception of top quark). The FN symmetry only  allows Yukawa-like couplings where at least one of the fermions is a messenger fermion: $\bar{f}\xi \phi$, $\bar{f}H\xi$ or $\bar{\xi}\xi\phi$. The messenger fieds are assumed to be much heavier than the SM particles and the flavon, and can be integrated out.  At the energy scales lower than the FN messenger mass the following effective operator can be constructed:
\begin{equation}\label{FN operator}
\mathcal{L}\supset c^{f}_{ij}\left(\frac{\phi}{\Lambda}\right)^{n^f_{ij}}\bar{\psi}_{L,i}^f H f_{R,j}+h.c.,
\end{equation}
where $c^f_{ij}$ is a dimensionless order-one number, $\Lambda$ is the mass scale of the messenger fermions, that  have been  integrated out, $\psi_{L,i}^f$ is $SU(2)_L$ fermion doublet and $f_{R,i}$  a $SU(2)_L$ fermion singlet, and $H$ is the SM Higgs doublet. The power $n^f_{ij}$ is determined by the FN charge conservation\footnote{It is understood that if $n^f_{ij}$ is negative the operator in Eq. (\ref{FN operator}) is replaced by $c^{f}_{ij}\left(\frac{\phi^\ast}{\Lambda}\right)^{-n^f_{ij}}\bar{\psi}_{L,i}^f H f_{R,j}$.}:
\begin{equation}
n^f_{ij}=-\frac{1}{q_{\phi}}(q_{\bar l,i}+q_{R,j}+q_{H}).
\end{equation}
 As the flavon acquires a VEV, this operator will give rise to the SM Yukawa-couplings. As we expand this operator around the vacuum, we obtain:
\begin{eqnarray}
&&\mathcal{L}\supset c^{f}_{ij}\left(\frac{\phi+\frac{v_\phi}{\sqrt{2}}}{\Lambda}\right)^{n^f_{ij}}\bar{\psi}_{L,i}^f (H+\langle H\rangle) f_{R,j}+h.c.\label{FN terms}\\
=&&c^{f}_{ij}\left(\frac{v_\phi}{\sqrt{2}\Lambda}\right)^{n^f_{ij}}\bar{f}_{L,i} f_{R,j}\frac{1}{\sqrt{2}}(h+v)
+n^f_{ij}c^{f}_{ij}\left(\frac{v_\phi}{\sqrt{2}\Lambda}\right)^{n^f_{ij}}\frac{v}{v_\phi}\bar{f}_{L,i} f_{R,j}~\phi+h.c.+\cdots,\nonumber
\end{eqnarray}
where we have ignored non-renormalizable terms in the second line. The first term in the second line  gives the SM Yukawa term and determines the Yukawa coupling of the Standard Model:
\begin{equation}
Y^f_{ij}=c^{f}_{ij}\left(\frac{v_\phi}{\sqrt{2}\Lambda}\right)^{n^f_{ij}}.
\end{equation}
Assuming that $v_\phi/(\sqrt{2}\Lambda)<1$, we obtain a hierarchical Yukawa matrix, by assigning larger FN charges to the lighter fermions compared to the heavier ones.  This way the charge assignment determines the hierarchy of the Yukawa couplings, in contrast to the Standard Model where the hierarchy is obtained  by fine-tuning the couplings themselves. 

The second term in the Eq. (\ref{FN terms}) describes  Yukawa-like interaction between the SM fermions and the flavon. The flavon coupling to the fermions is not proportional to the Yukawa matrix and is, therefore, flavour violating. This flavour violating coupling is inversely proportional to flavon VEV, which can suppress the flavour changing neutral currents mediated by the flavon given that the flavon VEV is large enough.

\subsection{The Froggatt-Nielsen mechanism  in the  331-framework}\label{FN in 331}

The vanilla FN-mechanism requires the introduction of new complex scalar field, the flavon, into the model. If we were to do it in 331-setting, would it make the already cumbersome scalar sector even more so. However,  the minimal  scalar sector we have introduced in the Eq.  (\ref{scalar triplets}) is compatible with the Froggatt-Nielsen mechanism as it is. The scalar triplets $\rho$ and $\chi$  have the same $U(1)_X$ charge and, therefore, the combination $\rho^{\dagger}\chi$ is a gauge singlet. According to Table \ref{FN charges of scalars}, this combination carries an overall FN charge and can  act as a flavon. Eq.   (\ref{vacuum}) shows the most general vacuum values for the scalar triplets. The most general vacuum structure is chosen in order the effective flavon to obtain a  non-zero  vacuum expectation value:  $\langle\rho^{\dagger}\chi\rangle=(v_2 u)/2$. 

The relevant effective operator for generating the Yukawa-couplings of the 331-model  is: 
\begin{equation}\label{FN331 operator}
\mathcal{L}\supset (c_s^{f})_{ij}\left(\frac{\rho^{\dagger}\chi}{\Lambda^2}\right)^{(n^s_f)_{ij}}\bar{\psi}_{L,i}^f S f_{R,j}+h.c.,
\end{equation}
where $(c_s^{f})_{ij}$  is a dimensionless order-one number,  $S$ denotes any of the three scalar triplets $\eta$, $\rho$ or $\chi$. The $\bar{\psi}_{L,i}^f$  and $f_{R,j}$ represent here the fermion triplets, anti-triplets and singlets that were introduced in Section \ref{fermion representations}. The $(n^s_f)_{ij}$ is determined by the FN charge assignment in the Table \ref{FN charges in 331}:
\begin{equation}
(n^s_f)_{ij}=\left[q(\bar{\psi}_{L,i}^f)+q(f_{R,j})+q(S)\right].
\end{equation}
If $(n^s_f)_{ij}$ were negative, we would  simply  include operator,
\be
\mathcal{L}\supset (c_s^{f})_{ij}\left(\frac{\chi^{\dagger}\rho}{\Lambda^2}\right)^{-(n^s_f)_{ij}}\bar{\psi}_{L,i}^f S f_{R,j}+h.c., 
\ee
 instead of the one in Eq. (\ref{FN331 operator}).

\begin{table}[h!]
\begin{center}
\begin{tabular}{|c|c|c|c|c|c|}
\hline
Particle   &  $\rho$ & $\chi $ & $(\psi^f_{L,i})^{c}$ & $f_{R,i}$ & $S$\\
\hline 
FN charge & 1  & 0 & $q(\bar{\psi}_{L,i}^f)$ & $q(f_{R,j})$ & $q_{S}$ \\
\hline
\end{tabular}
\end{center}
\caption{The FN charges of  fermions and the scalar fields.}
\label{FN charges in 331}
\end{table}

As the scalar triplets $\rho$ and $\chi$ acquire VEVs, the usual 331 Yukawa-terms are generated as effective couplings:
\begin{eqnarray}
&&\mathcal{L}\supset (c^f_s)_{ij}\left(\frac{(\rho+\langle\rho\rangle)^{\dagger}(\chi+\langle\chi\rangle)}{\Lambda^2}\right)^{(n^s_f)_{ij}}\bar{\psi}_{L,i}^f(S+\langle S\rangle) f_{R,j}+h.c.\label{331 FN}\\
&&=(y^f_s)_{ij}\bar{\psi}_{L,i}^f(S+\langle S\rangle) f_{R,j}
+(n^s_f)_{ij}(y^f_s)_{ij}\left[\frac{{\rho'}^{0\ast}}{v_2}+\frac{{\chi'}^0}{u}
+\frac{v_1 {\chi}^0 }{v_2 u}\right]\sqrt{2}\bar{\psi}_{L,i}^f\langle S\rangle f_{R,j}+h.c.+\cdots,\nonumber
\end{eqnarray} 
where we have kept only the renormalizable contributions. The Yukawa coupling is  defined  as:
\begin{equation}
(y^f_s)_{ij}=(c^f_s)_{ij}\left(\frac{v_2 u}{2\Lambda^2}\right)^{(n^s_f)_{ij}}\equiv 
(c^f_s)_{ij}\epsilon^{(n^s_f)_{ij}}.
\end{equation}
Just like in original FN model, the first term  in Eq.   (\ref{331 FN}) gives  the usual Yukawa terms of the model and the second term is a flavour violating part characteristic to Froggatt-Nielsen mechanism. Next we study the Yukawa  couplings  of our model in more detail.

\section{Yukawa couplings and fermion masses}\label{Yukawa couplings and fermion masses}
We now turn into the Yukawa couplings of the model. We classify the Yukawa couplings into our categories according to their contribution to the fermion mass. The charged lepton Yukawa couplings for example will contain the Yukawa couplings of charged leptons to neutral scalars, giving rise to the charged lepton masses, but also contain couplings  of charge scalars to neutrinos  and charged scalars. For completeness we include both the traditional 331 Yukawa couplings as well as the extra couplings originating from the Froggatt-Nielsen mechanism. 
\subsection{Charged lepton Yukawa couplings and masses}
The application of Eq. (\ref{331 FN}) to charged leptons  produces the following terms:
\begin{eqnarray}
&&\mathcal{L}_{\textrm{lepton}}
=y^{e}_{ij}\bar{L}_{L,i}\eta e_{R,j}+l_{ij}y^e_{ij}~\bar{e}_{L,i}e_{R,j}~\left[\frac{v'}{v_2}{\rho'}^{0\ast}+\frac{v'}{u}{\chi'}^{0}+\frac{v'v_1}{v_2 u}{\chi}^{0}\right]+h.c.,\label{lepton-yukawa}
\end{eqnarray}
where $i,j=1,2,3$. The Yukawa matrix is given by Froggatt-Nielsen as follows:
\begin{eqnarray}
&&y^e_{ij}=c^e_{ij}\epsilon^{q(\bar{L}_{L,i})+q(e_{R,j})+q(\eta)}
\equiv c^e_{ij}\epsilon^{l_{ij}}.
\label{lepton-epsilon}
\end{eqnarray}
 The first term in the Eq.   (\ref{lepton-yukawa}) is the usual Yukawa-coupling and the second term stems from the Froggatt-Nielsen mechanism.
All the charged lepton generations  are treated identically and there will be no flavour violation coming from the standard Yukawa coupling once the charge lepton mass matrix in diagonalized. 
 Notice that while the second term is flavour violating, it is heavily suppressed by the large $SU(3)_L$-breaking VEVs.   
The charged lepton masses originate from the first term in the Eq. (\ref{lepton-yukawa}) as the scalar  $\eta$ acquires VEV. The  charged lepton mass term is  simple  compared to the  masses of the rest of the fermions in the model: 
\begin{equation}
\mathcal{L}_{\textrm{lepton mass}}=m^e_{ij}\bar{e}_{L,i}e_{R,j}
+h.c.,
\end{equation}
where
\begin{eqnarray}
&&m^e_{ij}=\left(y^{e}_{ij}\frac{v'}{\sqrt{2}}\right).\nonumber
\end{eqnarray}
The charged lepton mass matrix  is proportional to the Yukawa matrix and there will be no flavour changing couplings in the standard Yukawa couplings.  The only flavour violation to the charged leptons is coming from the Froggatt-Nielsen mechanism but  is negligible as it is suppressed by $SU(3)_L\times U(1)_X$ breaking VEV.

\subsection{Neutrino Yukawa couplings and masses}
The neutrino Yukawa couplings do not originate from the operator in Eq. (\ref{331 FN}), but instead from the operator of the form,
\be
\mathcal{L}\supset (c^N_{\eta^\ast})_{ij}\left(\frac{\rho^{\dagger}\chi}{\Lambda^2}\right)^{(n^{\eta^\ast}_N)_{ij}}\epsilon_{\alpha\beta\gamma}\bar{L}_{L,i}^{\alpha}(L^{c}_{L,j})^{\beta}(\eta^{\ast})^{\gamma}+h.c.,
\ee

\begin{eqnarray}
&&\mathcal{L}_{\textrm{neutrino}}=e_{ij}\epsilon_{\alpha\beta\gamma}\bar{L}_{L,i}^{\alpha}(L^{c}_{L,j})^{\beta}(\eta^{\ast})^{\gamma}
\nonumber\\
&&+(n^{\eta}_N)_{ij}e_{ij}\epsilon_{\alpha\beta\gamma}\bar{L}_{L,i}^{\alpha}(L^{c}_{L,j})^{\beta}\langle\eta^\ast\rangle^\gamma
\sqrt{2}\left[\frac{{\rho'}^{0\ast}}{v_2}+\frac{{\chi'}^{0}}{u}+\frac{v_1}{v_2 u}{\chi}^{0}\right]+h.c.,\label{neutrino standard+FN}
\end{eqnarray}
where $i,j=1,2,3$ and the Yukawa couplings are 

\begin{eqnarray}
e_{ij}&=&(c^N_{\eta^\ast})_{ij}\epsilon^{(n^{\eta^\ast}_N)_{ij}},\nonumber
\end{eqnarray}
where
\begin{eqnarray}
 (n^{\eta^\ast}_N)_{ij}=q(L^c_{L,i})+q(L^c_{L,j})+q(\eta^\ast)=(n^{\eta^\ast}_N)_{ji}.\nonumber
\end{eqnarray}
The  coupling $e_{ij}$ is antisymmetric in its indices $i$ and $j$.
The first line in Eq.  (\ref{neutrino standard+FN}) contains the standard Yukawa interactions for the neutrino and the two last lines  contain the non-standard Yukawa interactions originating from the Froggatt-Nielsen mechanism. This flavour changing contribution is inversely proportional to heavy scales and is numerically negligible. 

The neutrino masses originate from the first term in Eq. (\ref{neutrino standard+FN}). The neutrino mass matrix will be antisymmetric which will result in one zero eigenvalue and two degenerate mass eigenvalues. Radiative corrections can lift the one  mass from zero and break the degeneracy of the other two as was demostrated in \cite{Valle:1983dk}. Generation of neutrino masses will be further discussed in \cite{Huitu:2019bvo},\cite{Huitu:2019mdr}.

\subsection{Up-type quark Yukawa couplings and masses}
The  first  generation of quarks is treated differently from the other two. This makes  the quark Yukawa couplings more complicated than  those of the  charged leptons. 
The up-type Yukawa couplings are: 
\begin{eqnarray}
&&\mathcal{L}_{up}
=\sum_{\gamma=1}^{4}(y^u_\rho)_{1\gamma}\bar{Q}'_{L,1}\rho ~u'_{R,\gamma}
+\sum_{\gamma=1}^{4}(y^u_\chi)_{1\gamma}\bar{Q}'_{L,1}\chi ~u'_{R,\gamma}
+\sum_{\alpha=2}^{3}\sum_{\gamma=1}^{4}(y^u_{\eta^{\ast}})_{\alpha\gamma}\bar{Q}'_{L,\alpha}\eta^{\ast} ~u'_{R,\gamma}\nonumber\\
&&+\left\{\sum_{\gamma=1}^{4}(n^{\rho}_{u})_{1\gamma}(y^u_{\rho})_{1\gamma}\bar{Q}'_{L,1}\langle\rho\rangle ~u'_{R,\gamma}
+\sum_{\gamma=1}^{4}(n^{\chi}_{u})_{1\gamma}(y^u_{\chi})_{1\gamma}\bar{Q}'_{L,1}\langle\chi\rangle ~u'_{R,\gamma}\right.\nonumber\\
&&+\left. \sum_{\alpha=2}^{3}\sum_{\gamma=1}^{4}(n^{\eta^\ast}_{u})_{\alpha\gamma}(y^u_{\eta^{\ast}})_{\alpha\gamma}\bar{Q}'_{L,\alpha}\langle\eta^{\ast}\rangle ~u'_{R,\gamma}\right\}
\sqrt{2}\left[\frac{{\rho'}^{0\ast}}{v_2}+\frac{{\chi'}^{0}}{u}+\frac{v_1{\chi}^{0}}{v_2 u}\right]+h.c.,\label{u-yukawa interaction}
\end{eqnarray}
where $u'_{R}=({u'}_{R,1},{u'}_{R,2},{u'}_{R,3},{U'}_{R})$. The first line contains the usual Yukawa-couplings of the up-type quarks and the last two lines contain the extra contribution from the FN-mechanism. The latter are numerically negligible and not shown in this section. The complete set is discussed in Appendix \ref{quark FN extra contribution}. 

The Yukawa couplings are given by the Froggatt-Nielsen mechanism as follows:
\begin{eqnarray}
&&(y^u_\rho)_{1\gamma}=(c^u_\rho)_{1\gamma}\epsilon^{q(\bar{Q}_{L,1})+q(u_{R,\gamma})+q(\rho)}
=(c^u_\rho)_{1\gamma}\epsilon^{(n^{\rho}_u)_{1\gamma}}\equiv (c^u_\rho)_{1\gamma}\epsilon^{(k^{\rho})_{1\gamma}},\nonumber\\
&&(y^u_{\eta^{\ast}})_{\alpha\gamma}=(c^u_{\eta^{\ast}})_{\alpha\gamma}\epsilon^{q(\bar{Q}_{L,\alpha})+q(u_{R,\gamma})+q(\eta^{\ast})}
=(c^u_{\eta^{\ast}})_{1\gamma}\epsilon^{(n^{\eta^{\ast}}_u)_{\alpha\gamma}}\equiv
(c^u_{\eta^{\ast}})_{1\gamma}\epsilon^{(k^{\eta^{\ast}})_{\alpha\gamma}},\nonumber\\
&&(y^u_\chi)_{1\gamma}=(c^u_\chi)_{1\gamma}\epsilon^{q(\bar{Q}_{L,1})+q(u_{R,\gamma})+q(\chi)}
=(c^u_\chi)_{1\gamma}\epsilon^{(n^{\chi}_u)_{1\gamma}}
\equiv(c^u_\chi)_{1\gamma}\epsilon^{(k^{\chi})_{1\gamma}},\label{up-epsilon}
\end{eqnarray}
where $\alpha=2,3$ and $\gamma=1,2,3,4$.

More explicitly, the Lagrangian for neutral scalars coupling to up-type quarks is,
\begin{eqnarray}
\mathcal{L}_{up-type} &=& \bar{u}'_{L} \widetilde{\Gamma}^{u}_{\rho} u'_{R}~\rho^0+\bar{u}'_{L} \widetilde{\Gamma}^{u}_{\rho'} u'_{R}~{\rho'}^{0}+\bar{u}'_{L} \widetilde{\Gamma}^{u}_{\chi} u'_{R}~{\chi}^{0}
+\bar{u}'_{L} \widetilde{\Gamma}^{u}_{\chi'} u'_{R}~{\chi'}^{0}
+\bar{u}'_{L} \widetilde{\Gamma}^{u}_{\eta} u'_{R}~{\eta}^{0\ast},\label{up type quark coupling to neutral scalars}
\end{eqnarray}
where,
\begin{eqnarray}
&&\widetilde{\Gamma}^{u}_{\rho}=\left(\begin{array}{cccc}
(y^{u}_{\rho})_{11} & (y^{u}_{\rho})_{12} & (y^{u}_{\rho})_{13} & (y^{u}_{\rho})_{14} \\ 
0 & 0 & 0 & 0 \\
0 & 0 & 0 & 0 \\ 
0 & 0 & 0 & 0 \\
\end{array}\right),\quad 
\widetilde{\Gamma}^{u}_{\rho'}=\left(\begin{array}{cccc}
0 & 0 & 0 & 0 \\
0 & 0 & 0 & 0 \\
0 & 0 & 0 & 0 \\ 
(y^{u}_{\rho})_{11} & (y^{u}_{\rho})_{12} & (y^{u}_{\rho})_{13} & (y^{u}_{\rho})_{14} \\ 
\end{array}\right),\nonumber\\
&&\widetilde{\Gamma}^{u}_{\chi}=\left(\begin{array}{cccc}
(y^{u}_{\chi})_{11} & (y^{u}_{\chi})_{12} & (y^{u}_{\chi})_{13} & (y^{u}_{\chi})_{14} \\ 
0 & 0 & 0 & 0 \\
0 & 0 & 0 & 0 \\
0 & 0 & 0 & 0 \\
\end{array}\right),\quad
\widetilde{\Gamma}^{u}_{\chi'}=\left(\begin{array}{cccc}
0 & 0 & 0 & 0 \\
0 & 0 & 0 & 0 \\
0 & 0 & 0 & 0 \\ 
(y^{u}_{\chi})_{11} & (y^{u}_{\chi})_{12} & (y^{u}_{\chi})_{13} & (y^{u}_{\chi})_{14} \\ 
\end{array}\right),\nonumber\\
&&\widetilde{\Gamma}^{u}_{\eta}=\left(\begin{array}{cccc}
0 & 0 & 0 & 0 \\ 
-(y^{u}_{\eta^\ast})_{21}  & -(y^{u}_{\eta^\ast})_{22} & -(y^{u}_{\eta^\ast})_{23} & -(y^{u}_{\eta^\ast})_{24} \\ 
-(y^{u}_{\eta^\ast})_{31}  & -(y^{u}_{\eta^\ast})_{32} & -(y^{u}_{\eta^\ast})_{33} & -(y^{u}_{\eta^\ast})_{34} \\ 
0 & 0 & 0 & 0 \\ 
\end{array}\right).\nonumber
\end{eqnarray}

The up-type quark masses are generated by the terms in the Eq. (\ref{up type quark coupling to neutral scalars}).
The up-quark mass matrix in the basis,
\begin{equation}
\mathcal{L}_{up-mass}=\bar{u}'_{L} m_u u'_R+h.c.,
\end{equation}
is,
\begin{eqnarray}
\resizebox{1.0 \textwidth}{!} 
{
$ 
m_u=\frac{1}{\sqrt{2}}\left(
\begin{array}{cccc}
v_1 (y^{u}_{\rho})_{11} &
v_1  (y^{u}_{\rho})_{12} &
v_1  (y^{u}_{\rho})_{13} &
 v_1(y^{u}_{\rho})_{14}\\
-v' (y^{u}_{\eta^*})_{21}&
 -v' (y^{u}_{\eta^*})_{22}   & 
-v' (y^{u}_{\eta^*})_{23} &
 -v' (y^{u}_{\eta^*})_{24} \\
-v' (y^{u}_{\eta^*})_{31}  &
 -v'(y^{u}_{\eta^*})_{32}   &
 -v' (y^{u}_{\eta^*})_{33} &
 -v' (y^{u}_{\eta^*})_{34}    \\
v_2 (y^{u}_{\rho})_{11}+u(y^{u}_{\chi})_{11}   &
 v_2(y^{u}_{\rho})_{12}+u(y^{u}_{\chi})_{12}   & 
v_2(y^{u}_{\rho})_{13}+u(y^{u}_{\chi})_{13}&
 v_2 (y^{u}_{\rho})_{14}+u(y^{u}_{\chi})_{14}  \\
\end{array}
\right).\nonumber
$
}
\end{eqnarray}
In the  331-setting the hierachy of the quark mass matrix elements is determined by the FN-charge assignment and different VEVs, in contrast to traditional FN-mechanism, where the hierarchy is set solely by the charge assignment. The $SU(3)_L$ breaking VEVs $u$ and $v_2$ are assumed to be much larger than those of the  $SU(2)_L$ breaking. This greatly affects the hierarchy.
 We rewrite the mass matrix elements so that the hierarchy is more transparent:

\begin{eqnarray}
&&(m_u)_{1\gamma}=\frac{v'}{\sqrt{2}} \left[\frac{v_1}{v'}(c_{\rho}^{u})_{1\gamma}\right]\epsilon^{a_1^{u}+q(u_{R,\gamma})},\nonumber\\
&&(m_u)_{2\gamma}= \frac{v'}{\sqrt{2}} \left[-(c_{\eta^*}^{u})_{2\gamma}\right]\epsilon^{a_2^{u}+q(u_{R,\gamma})},\nonumber\\
&&(m_u)_{3\gamma}=\frac{v'}{\sqrt{2}} \left[ -(c_{\eta^*}^{u})_{3\gamma}\right]\epsilon^{a_3^{u}+q(u_{R,\gamma})},\nonumber\\
&&(m_u)_{4\gamma}=\frac{v'}{\sqrt{2}} \left[(c_{\rho}^{u})_{1\gamma}\epsilon^{q(\rho)-q(\chi)}+(c_{\chi}^{u})_{1\gamma}\epsilon^{(\log \epsilon)^{-1}\log(u/v_2)}\right]\epsilon^{a_4^{u}+q(u_{R,\gamma})},\nonumber
\end{eqnarray}
where the  quantities in square brackets are order-one numbers, and therefore the hierarchy is completely set by the powers of $\epsilon$. The  $a^{u}_{\gamma}$ are:
\begin{eqnarray}
&&a^{u}_1= q(\bar{Q}_{L,1})+q(\rho),\label{effective charges up}\\
&&a^{u}_2= q(\bar{Q}_{L,2})+q(\eta^*),\nonumber\\
&&a^{u}_3= q(\bar{Q}_{L,3})+q(\eta^*),\nonumber\\
&&a^{u}_4= (\log \epsilon)^{-1}\log\left(\frac{v_2}{v'}\right)+q(\bar{Q}_{L,1})+q(\chi).\nonumber
\end{eqnarray}
The difference between two symmetry breaking scales manifests itself as effective left-handed charge.

They are effective left-handed charges that are analogous to FN  charges of left-handed fermion doublets in the original FN mechanism.  By writing the  matrix elements in this way, the hierarchy is most transparent and the textures of the  diagonalization matrices are  easily obtained.

\subsection{Down-type quark Yukawa couplings and masses}
 The down-type quark Yukawa couplings are  written  similarly to the up-type couplings  separating the first generation from the rest.
The down-type quark Yukawa-couplings are, 
\begin{eqnarray}
&&\mathcal{L}_{down}
=\sum_{\gamma=1}^{5}(y^d_{\eta})_{1\gamma}\bar{Q}'_{L,1}\eta ~d'_{R,\gamma}
+\sum_{\alpha=2}^{3}\sum_{\gamma=1}^{5}(y^d_{\rho^{\ast}})_{\alpha\gamma}\bar{Q}'_{L,\alpha}\rho^{\ast} ~d'_{R,\gamma}
+\sum_{\alpha=2}^{3}\sum_{\gamma=1}^{5}(y^d_{\chi^{\ast}})_{\alpha\gamma}\bar{Q}'_{L,\alpha}\chi^{\ast} ~d'_{R,\gamma}\nonumber\\
&&+\left\{\sum_{\gamma=1}^{5}(n^{\eta}_{d})_{1\gamma}(y^d_{\eta})_{1\gamma}\bar{Q}'_{L,1}\langle\eta\rangle ~d'_{R,\gamma}
+\sum_{\alpha=2}^{3}\sum_{\gamma=1}^{5}(n^{\rho^{\ast}}_{d})_{\alpha\gamma}(y^d_{\rho^{\ast}})_{\alpha\gamma}\bar{Q}'_{L,\alpha}\langle\rho^{\ast}\rangle ~d'_{R,\gamma}\right.\nonumber\\
&&+\left. \sum_{\alpha=2}^{3}\sum_{\gamma=1}^{5}(n^{\chi^\ast}_{d})_{\alpha\gamma}(y^d_{\chi^{\ast}})_{\alpha\gamma}\bar{Q}'_{L,\alpha}\langle\chi^{\ast}\rangle ~d'_{R,\gamma}\right\}
\sqrt{2}\left[\frac{{\rho'}^{0\ast}}{v_2}+\frac{{\chi'}^{0}}{u}+\frac{v_1{\chi}^{0}}{v_2 u}\right]+h.c.,\label{d-yukawa interaction}
\end{eqnarray}
where $d'_{R}=({d'}_{R,1},{d'}_{R,2},{d'}_{R,3},{D'}_{R,1},{D'}_{R,2})$. The first line contains the standard 331 Yukawa-couplings of the down-type quarks and the last two lines contain the extra contribution from the FN-mechanism. The latter are numerically negligible and not shown in this section. The complete set is discussed in Appendix \ref{quark FN extra contribution}. 

The Yukawa couplings are given by the Froggatt-Nielsen mechanism as follows:
\begin{eqnarray}
&&(y^d_\eta)_{1\gamma}=(c^d_\eta)_{1\gamma}\epsilon^{q(\bar{Q}_{L,1})+q(d_{R,\gamma})+q(\eta)}
=(c^d_\eta)_{1\gamma}\epsilon^{(n^{\eta}_d)_{1\gamma}}
\equiv(c^d_\eta)_{1\gamma}\epsilon^{(h^{\eta})_{1\gamma}},\nonumber\\
&&(y^d_{\rho^{\ast}})_{\alpha\gamma}=(c^d_{\rho^{\ast}})_{\alpha\gamma}\epsilon^{q(\bar{Q}_{L,\alpha})+q(d_{R,\gamma})+q(\rho^{\ast})}
=(c^d_{\rho^{\ast}})_{1\gamma}\epsilon^{(n^{\rho^\ast}_d)_{\alpha\gamma}}\equiv
(c^d_{\rho^{\ast}})_{1\gamma}\epsilon^{(h^{\rho^\ast})_{\alpha\gamma}},\nonumber\\
&&(y^d_{\chi^{\ast}})_{\alpha\gamma}=(c^d_{\chi^{\ast}})_{\alpha\gamma}\epsilon^{q(\bar{Q}_{L,\alpha})+q(d_{R,\gamma})+q(\chi^{\ast})}
=(c^d_{\chi^{\ast}})_{1\gamma}\epsilon^{(n^{\chi^\ast}_d)_{\alpha\gamma}}\equiv 
(c^d_{\chi^{\ast}})_{1\gamma}\epsilon^{(h^{\chi^\ast})_{\alpha\gamma}},\label{down-epsilon}
\end{eqnarray}
where $\alpha=2,3$ and $\gamma=1,2,3,4,5$.

Explicit terms of the Lagrangian for neutral scalars coupling to down-type quarks  are,
\begin{eqnarray}
\mathcal{L}_{down-type} = \bar{d}'_{L} \widetilde{\Gamma}^{d}_{\eta} d'_{R}~\eta^0+\bar{d}'_{L} \widetilde{\Gamma}^{d}_{\rho} d'_{R}~\rho^{0\ast}+\bar{d}'_{L} \widetilde{\Gamma}^{d}_{\rho'} d'_{R}~{\rho'}^{0\ast}+\bar{d}'_{L} \widetilde{\Gamma}^{d}_{\chi} d'_{R}~\chi^{0\ast}+\bar{d}'_{L} \widetilde{\Gamma}^{d}_{\chi'} d'_{R}~{\chi'}^{0\ast},\label{down type quark coupling to neutral scalars}
\end{eqnarray}
where,
\begin{eqnarray}
&&\widetilde{\Gamma}^{d}_{\eta}=\left(\begin{array}{ccccc}
y_{11}^{d} & y_{12}^{d} & y_{13}^{d} & y_{14}^{d} & y_{15}^{d}\\
0 & 0 & 0 & 0 & 0\\ 
0 & 0 & 0 & 0 & 0\\ 
0 & 0 & 0 & 0 & 0\\ 
0 & 0 & 0 & 0 & 0\\ 
\end{array}\right),\nonumber\\
&&\widetilde{\Gamma}^{d}_{\rho}=\left(\begin{array}{ccccc}
0 & 0 & 0 & 0 & 0\\ 
y_{21}^{d} & y_{22}^{d} & y_{23}^{d} & y_{24}^{d} & y_{25}^{d}\\ 
y_{31}^{d} & y_{32}^{d} & y_{33}^{d} & y_{34}^{d} & y_{35}^{d}\\ 
0 & 0 & 0 & 0 & 0\\ 
0 & 0 & 0 & 0 & 0\\ 
\end{array}\right),\quad 
\widetilde{\Gamma}^{d}_{\rho'}=\left(\begin{array}{ccccc}
0 & 0 & 0 & 0 & 0\\
0 & 0 & 0 & 0 & 0\\ 
0 & 0 & 0 & 0 & 0\\ 
y_{21}^{d} & y_{22}^{d} & y_{23}^{d} & y_{24}^{d} & y_{25}^{d}\\ 
y_{31}^{d} & y_{32}^{d} & y_{33}^{d} & y_{34}^{d} & y_{35}^{d}\\  
\end{array}\right),\nonumber\\
&&\widetilde{\Gamma}^{d}_{\chi}=\left(\begin{array}{ccccc}
0 & 0 & 0 & 0 & 0\\ 
{y'}_{21}^{d} & {y'}_{22}^{d} & {y'}_{23}^{d} & {y'}_{24}^{d} & {y'}_{25}^{d}\\ 
{y'}_{31}^{d} & {y'}_{32}^{d} & {y'}_{33}^{d} & {y'}_{34}^{d} & {y'}_{35}^{d}\\ 
0 & 0 & 0 & 0 & 0\\ 
0 & 0 & 0 & 0 & 0\\
\end{array}\right),\quad
\widetilde{\Gamma}^{d}_{\chi'}=\left(\begin{array}{ccccc}
0 & 0 & 0 & 0 & 0\\ 
0 & 0 & 0 & 0 & 0\\ 
0 & 0 & 0 & 0 & 0\\ 
{y'}_{21}^{d} & {y'}_{22}^{d} & {y'}_{23}^{d} & {y'}_{24}^{d} & {y'}_{25}^{d}\\ 
{y'}_{31}^{d} & {y'}_{32}^{d} & {y'}_{33}^{d} & {y'}_{34}^{d} & {y'}_{35}^{d}\\
\end{array}\right).\nonumber
\end{eqnarray}

The down-type quark masses are generated by the terms in  Eq. (\ref{down type quark coupling to neutral scalars}).
The down-quark mass matrix in the basis:
\begin{equation}
\mathcal{L}_{down-mass}=\bar{d}'_{L} m_d d'_R+h.c.,
\end{equation}
is
\begin{eqnarray}
\resizebox{1.0 \textwidth}{!} 
{
$ 
m_d=
  \frac{1}{\sqrt{2}}\left(
\begin{array}{ccccc}
v'(y^{d}_{\eta})_{11} & 
v'(y^{d}_{\eta})_{12} &
v' (y^{d}_{\eta})_{13} & 
v'(y^{d}_{\eta})_{14} &
v' (y^{d}_{\eta})_{15} \\
v_1 (y^{d}_{\rho^*})_{21}&
v_1 (y^{d}_{\rho^*})_{22}  &
v_1 (y^{d}_{\rho^*})_{23}& 
v_1 (y^{d}_{\rho^*})_{24} &
v_1  (y^{d}_{\rho^*})_{25}  \\
v_1 (y^{d}_{\rho^*})_{31}  & 
v_1 (y^{d}_{\rho^*})_{32}  &
v_1  (y^{d}_{\rho^*})_{33}&
v_1  (y^{d}_{\rho^*})_{34} &
 v_1 (y^{d}_{\rho^*})_{35}  \\
v_2 (y^{d}_{\rho^*})_{21}+u(y^{d}_{\chi^*})_{21}  &
v_2 (y^{d}_{\rho^*})_{22}+u(y^{d}_{\chi^*})_{22}  &
 v_2 (y^{d}_{\rho^*})_{23}+u(y^{d}_{\chi^*})_{23}&
 v_2 (y^{d}_{\rho^*})_{24}+u(y^{d}_{\chi^*})_{24} &
v_2 (y^{d}_{\rho^*})_{25}+u(y^{d}_{\chi^*})_{25} \\
v_2 (y^{d}_{\rho^*})_{31}+u(y^{d}_{\chi^*})_{31}   &
v_2 (y^{d}_{\rho^*})_{32}+u(y^{d}_{\chi^*})_{32}   & 
v_2 (y^{d}_{\rho^*})_{33}+u(y^{d}_{\chi^*})_{33}& 
v_2 (y^{d}_{\rho^*})_{34}+u(y^{d}_{\chi^*})_{34}&
v_2 (y^{d}_{\rho^*})_{35}+u(y^{d}_{\chi^*})_{35}  \\
\end{array}
\right)\nonumber
$
}
\end{eqnarray}
We rewrite the down-type mass matrix elements in more transparent way in analogy to the up-type quark masses as:
\begin{eqnarray}
&&(m_d)_{1\gamma}=\frac{v'}{\sqrt{2}} \left[(c_{\eta}^{d})_{1\gamma}\right]\epsilon^{a_1^{d}+q(d_{R,\gamma})},\nonumber\\
&&(m_d)_{2\gamma}= \frac{v'}{\sqrt{2}} \left[\frac{v_1}{v'}(c_{\rho^*}^{d})_{2\gamma}\right]\epsilon^{a_2^{d}+q(d_{R,\gamma})},\nonumber\\
&&(m_d)_{3\gamma}= \frac{v'}{\sqrt{2}} \left[\frac{v_1}{v'}(c_{\rho^*}^{d})_{3\gamma}\right]\epsilon^{a_3^{d}+q(d_{R,\gamma})},\nonumber\\
&&(m_d)_{4\gamma}=\frac{v'}{\sqrt{2}} \left[(c_{\rho^*}^{d})_{2\gamma}+(c_{\chi^*}^{d})_{2\gamma}\epsilon^{(\log \epsilon)^{-1}\log(u/v_2)+q(\chi^*)-q(\rho^*)}\right]\epsilon^{a_4^{d}+q(d_{R,\gamma})},\nonumber\\
&&(m_d)_{5\gamma}=\frac{v'}{\sqrt{2}} \left[(c_{\rho^*}^{d})_{3\gamma}+(c_{\chi^*}^{d})_{3\gamma}\epsilon^{(\log \epsilon)^{-1}\log(u/v_2)+q(\chi^*)-q(\rho^*)}\right]\epsilon^{a_5^{d}+q(d_{R,\gamma})},\nonumber
\end{eqnarray}
where the  quantities in square brackets are order-one numbers, and therefore the hierarchy is completely set by the powers of $\epsilon$. The  $a^{d}_{\gamma}$ are :
\begin{eqnarray}
&&a^{d}_1= q(\bar{Q}_{L,1})+q(\eta),\label{effective charges down}\\
&&a^{d}_2= q(\bar{Q}_{L,2})+q(\rho^*),\nonumber\\
&&a^{d}_3= q(\bar{Q}_{L,3})+q(\rho^*),\nonumber\\
&&a^{d}_4= (\log \epsilon)^{-1}\log\left(\frac{v_2}{v'}\right)+q(\bar{Q}_{L,2})+q(\rho^*),\nonumber\\
&&a^{d}_5= (\log \epsilon)^{-1}\log\left(\frac{v_2}{v'}\right)+q(\bar{Q}_{L,3})+q(\rho^*).\nonumber
\end{eqnarray}

\section{The scalar FCNCs}\label{scalar FCNC}

Experimentally the flavour changing neutral currents are known to be suppressed. The Standard Model neatly explains this by assigning all the fermion generations to the same representation. In this way the neutral gauge bosons couplings to fermions are flavour diagonal and the loop-induced neutral currents mediated by the $W_\mu^{\pm}$ acquire additional suppression  \cite{Glashow:1970gm}. There is also only one Higgs-doublet coupling to each fermion type, so the fermion mass matrices are proportional to the Higgs Yukawa coupling, and so the Yukawa couplings are diagonalized simultaneously with the mass matrices, resulting to flavour conservation in the Higgs sector.

The situation in the 331-models is different. 
The anomaly cancellation in the traditional 331-model  requires  to treat fermion generations differently. This inevitably leads to flavour changing processes in the neutral scalars. The lepton generations  in our model are assigned into same representation and there will be no flavour changing couplings for the charged leptons in the scalar or neutral gauge boson sectors other than the   FN contributions and FCNCs are tiny as discussed earlier. 

We  have chosen to treat the first quark  generation differently from the others. 
 The quarks couple to more than one scalar multiplet giving rise to flavour change in the scalar sector. In this section we are concentrating to the flavour changing neutral currents of scalars. 
Also the   neutral gauge bosons $Z_\mu^0$, ${Z'}^0_\mu$ and $X^0_\mu$ mediate the quark FCNCs at tree-level. 
The GIM mechanism works for the $Z_\mu^0$-boson in a sense that all the coupligs of the SM quarks are identical to the $Z_\mu^0$-boson, only the coupling to the exotic quarks differ \cite{Montero:1992jk},\cite{Ozer:1996jc}. This will induce  flavour violating couplings between SM quarks and the exotic quarks, which is not experimentally that  constrained as the exotic quarks appear only at loop-order. The flavour violating coupling between SM quarks remains small and the $Z_\mu^0$-boson does not induce large FCNCs. The GIM mechanism does not take place  in the case of ${Z'}_\mu$ and $X^0_\mu$, but the FCNC effects mediated by them are suppresed due to their large mass.  We have numerically checked that the neutral gauge boson mediation passes the FCNC bounds. The neutral currents in the context of $\beta=\pm 1/\sqrt{3}$ have been studied in \cite{Singer:1980sw},  \cite{Montero:1992jk}, \cite{Hoang:1995vq},  \cite{Hoang:1996gi}, \cite{Pleitez:1994pu}, \cite{Dong:2006mg},  \cite{Ozer:1996jc}, \cite{Ozer:1995xi}, \cite{Dong:2008ya}.


Our FN331-model has physical six neutral  scalars, $h$, $H_1$, $H_2$, $H_3$, $A_1$ and $A_2$,  of which $H_1$, $H_2$, $H_3$ and $A_1$ are heavy as discussed in Section \ref{scalar masses}  and the FCNCs mediated by them are suppressed by their masses. The two remaining particles are the Higgs $h$ and the pseudo-Goldstone boson $A_2$. The Higgs is presumably the lightest scalar and, therefore, the flavour changing effects mediated by it will be the most devastating. 
The mass of the  pseudo-Goldstone boson $A_2$ is  a free parameter in our model. We assume that $m_{A_2}>> m_h$, so that it can be ignored as a  dangerous mediator of FCNCs.

\subsection{FCNC's mediated by Higgs}
To see the flavour-violating couplings of the  light physical neutral scalars, we write  quark Yukawa-interactions  Eq. (\ref{up type quark coupling to neutral scalars}) and Eq. (\ref{down type quark coupling to neutral scalars})    in terms of the physical Higgs, using Eq. (\ref{CP-even rotation}),
\begin{eqnarray}
\mathcal{L}_{\textrm{quark-Higgs}}&=&\frac{1}{\sqrt{2}}\bar{u}'_{L}({\Gamma'}^u_{h})u'_{R}~h+\frac{1}{\sqrt{2}}\bar{d}'_{L}({\Gamma'}^d_{h})d'_{R}~h+h.c.\label{higgs yukawa couplings}\\
&=& \frac{1}{\sqrt{2}}\bar{u}_{L}U_L^u({\Gamma'}^u_{h})U_{R}^{u\dagger}u_{R}~h+\frac{1}{\sqrt{2}}\bar{d}_{L}U_{L}^d({\Gamma'}^d_{h})U_{R}^{d\dagger}d'_{R}~h+h.c.,\nonumber
\end{eqnarray}
where the primes denote  gauge eigenstates  and  the coupling matrices are:
\begin{equation}
\resizebox{1.0 \textwidth}{!} 
{
$ 
{\Gamma'}^{u}_h=\left(
\begin{array}{cccc}
U^{H}_{22}(y^{u}_{\rho})_{11}+U^{H}_{32}(y^{u}_{\chi})_{11}  &
U^{H}_{22}(y^{u}_{\rho})_{12}+U^{H}_{32}(y^{u}_{\chi})_{12} & 
U^{H}_{22}(y^{u}_{\rho})_{13}+U^{H}_{32}(y^{u}_{\chi})_{13} & 
U^{H}_{22}(y^{u}_{\rho})_{14}+U^{H}_{32}(y^{u}_{\chi})_{14}\\
-U^{H}_{12}(y^{u}_{\eta^*})_{21} &
 -U^{H}_{12}(y^{u}_{\eta^*})_{22}   &
-U^{H}_{12} (y^{u}_{\eta^*})_{23} &
-U^{H}_{12} (y^{u}_{\eta^*})^{u}   \\
-U^{H}_{12}(y^{u}_{\eta^*})_{31}  &
-U^{H}_{12}(y^{u}_{\eta^*})_{32}   &
 -U^{H}_{12}(y^{u}_{\eta^*})_{33} &
-U^{H}_{12}(y^{u}_{\eta^*})^{u}    \\
U^{H}_{42}(y^{u}_{\rho})_{11}+U^{H}_{52}(y^{u}_{\chi})_{11}   &
 U^{H}_{42}(y^{u}_{\rho})_{12}+U^{H}_{52}(y^{u}_{\chi})_{12}   &
 U^{H}_{42}(y^{u}_{\rho})_{13}+U^{H}_{52}(y^{u}_{\chi})_{13} &
 U^{H}_{42}(y^{u}_{\rho})_{14}+U^{H}_{52}(y^{u}_{\chi})_{14}, \nonumber
\end{array}
\right),
$
}
\end{equation}
and,
\begin{equation}
\resizebox{1.0 \textwidth}{!} 
{
$ 
{\Gamma'}^d_h=\left(
\begin{array}{ccccc}
 U^{H}_{12}(y^{d}_{\eta})_{11} &  
U^{H}_{12}(y^{d}_{\eta})_{12} &
 U^{H}_{12} (y^{d}_{\eta})_{13} &
 U^{H}_{12} (y^{d}_{\eta})^{d} & 
U^{H}_{12}(y^{d}_{\eta})^{d} \\
U^{H}_{22}(y^{d}_{\rho^*})_{21}+U^{H}_{32}(y^{d}_{\chi^*})_{21} &
 U^{H}_{22}(y^{d}_{\rho^*})_{22}+U^{H}_{32}(y^{d}_{\chi^*})_{22} &
 U^{H}_{22}(y^{d}_{\rho^*})_{23}+U^{H}_{32}(y^{d}_{\chi^*})_{23} &
U^{H}_{22}(y^{d}_{\rho^*})_{24}+U^{H}_{32}(y^{d}_{\chi^*})_{24} &
 U^{H}_{22}(y^{d}_{\rho^*})_{25}+U^{H}_{32}(y^{d}_{\chi^*})_{25}  \\
U^{H}_{22}(y^{d}_{\rho^*})_{31}+U^{H}_{32}(y^{d}_{\chi^*})_{31} & 
U^{H}_{22}(y^{d}_{\rho^*})_{32}+U^{H}_{32}(y^{d}_{\chi^*})_{32} &
 U^{H}_{22}(y^{d}_{\rho^*})_{33}+U^{H}_{32}(y^{d}_{\chi^*})_{33} &
U^{H}_{22}(y^{d}_{\rho^*})_{34}+U^{H}_{32}(y^{d}_{\chi^*})_{34} & 
U^{H}_{22}(y^{d}_{\rho^*})_{35}+U^{H}_{32}(y^{d}_{\chi^*})_{35}  \\
U^{H}_{42}(y^{d}_{\rho^*})_{21}+U^{H}_{52}(y^{d}_{\chi^*})_{21} &
 U^{H}_{42}(y^{d}_{\rho^*})_{22}+U^{H}_{42}(y^{d}_{\chi^*})_{22} &
 U^{H}_{42}(y^{d}_{\rho^*})_{23}+U^{H}_{52}(y^{d}_{\chi^*})_{23} &
U^{H}_{42}(y^{d}_{\rho^*})_{24}+U^{H}_{52}(y^{d}_{\chi^*})_{24} & 
U^{H}_{42}(y^{d}_{\rho^*})_{25}+U^{H}_{52}(y^{d}_{\chi^*})_{25} \\
U^{H}_{42}(y^{d}_{\rho^*})_{31}+U^{H}_{52}(y^{d}_{\chi^*})_{31} &
 U^{H}_{42}(y^{d}_{\rho^*})_{32}+U^{H}_{42}(y^{d}_{\chi^*})_{32} &
 U^{H}_{42}(y^{d}_{\rho^*})_{33}+U^{H}_{52}(y^{d}_{\chi^*})_{33} &
U^{H}_{42}(y^{d}_{\rho^*})_{34}+U^{H}_{52}(y^{d}_{\chi^*})_{34} &
 U^{H}_{42}(y^{d}_{\rho^*})_{35}+U^{H}_{52}(y^{d}_{\chi^*})_{35}  \nonumber
\end{array}
\right).
$
}
\end{equation}
In Eq. (\ref{higgs yukawa couplings}) terms inversely proportional to the large scale coming from the FN mechanism  have been dropped. 
The ${\Gamma'}^{u}_h$ and ${\Gamma'}^{d}_h$ represent the Yukawa  interactions of 331-model.  The contributions coming from the FN mechanism  are discussed in the Appendix \ref{quark FN extra contribution}.

The physical Yukawa couplings for quarks are obtained once the physical quark fields are introduced. The physical Yukawa couplings $\Gamma^u_h=U_L^u{\Gamma'}^u_{h}U_{R}^{u\dagger}$ and  $\Gamma^d_h=U_L^d{\Gamma'}^d_{h}U_{R}^{d\dagger}$ can be written    as:
\begin{eqnarray}
(\Gamma^u_h)_{ij} & = & \sqrt{2}U^{H}_{12}\frac{m_j}{v'}\delta_{ij}\label{fcnc up}\\
&&+\sqrt{2}\left(\frac{U^{H}_{22}}{v_1}-\frac{U^{H}_{12}}{v'}-U^{H}_{32}\frac{v_2}{v_1 u}\right)m_j (U_L^u)_{i1}(U_L^{u\dagger})_{1j}\nonumber\\
&&+\sqrt{2}\left(\frac{U^{H}_{52}}{u}-\frac{U^{H}_{12}}{v'}\right)m_j (U_L^u)_{i4}(U_L^{u\dagger})_{4j}\nonumber\\
&&
+\sqrt{2}\left[\frac{U^{H}_{32}}{u}\right]m_j (U_L^u)_{i1} (U_L^{u\dagger})_{4j}
+\sqrt{2}\left[\frac{U^{H}_{42}}{v_1}-\frac{v_2}{v_1 u}U^{H}_{52}\right]m_j (U_L^u)_{i4} (U_L^{u\dagger})_{1j},
\nonumber
\end{eqnarray}
and,
\begin{eqnarray}
(\Gamma^d_h)_{ij} & = & \sqrt{2}U^{H}_{12}\frac{m_j}{v'}\delta_{ij}\label{fcnc down}\\
&&+\sqrt{2}\left(\frac{U^{H}_{22}}{v_1}-\frac{U^{H}_{12}}{v'}-U^{H}_{32}\frac{v_2}{v_1 u}\right)m_j \left[(U_L^u)_{i2}(U_L^{u\dagger})_{2j}+(U_L^u)_{i3}(U_L^{u\dagger})_{3j}\right]
\nonumber\\
&&+\sqrt{2}\left[\frac{U^{H}_{52}}{u}-\frac{U^{H}_{12}}{v'}\right]m_j \left[(U_L^d)_{i4}(U_L^{d\dagger})_{4j}+ (U_L^d)_{i5}(U_L^{d\dagger})_{5j}\right]
\nonumber\\
&&+\sqrt{2}\left[\frac{U^{H}_{32}}{u}\right]m_j \left[(U_L^{d})_{i2}(U_L^{d\dagger})_{4j}
+(U_L^{d})_{i3}(U_L^{d\dagger})_{5j}\right]
\nonumber\\
&&
+\sqrt{2}\left[\frac{U^{H}_{42}}{v_1}-\frac{v_2}{v_1 u}U^{H}_{52}\right] m_j \left[(U_L^{d})_{i4}(U_L^{d\dagger})_{2j}
+(U_L^{d})_{i5}(U_L^{d\dagger})_{3j}\right].
\nonumber
\end{eqnarray}
The first lines in Eq.  (\ref{fcnc up}) and  Eq. (\ref{fcnc down}) are contributions to the diagonal elements of physical Yukawa couplings of quarks. The rest are giving all the off-diagonal contributions, but do contribute to the diagonal elements as well. The off-diagonal terms have to be sufficiently suppressed, in order to avoid large flavour changing neutral currents.  According to Eq. (\ref{higgs eigenvector approximation}),  cancellations take  place among the  Higgs eigenvector components.  The leading contribution from Higgs eigenvector components  cancels in the second lines of Eq. (\ref{fcnc up}) and Eq.  (\ref{fcnc down}). By  using the Eq.   (\ref{higgs eigenvector approximation}), the Eqs.  (\ref{fcnc up}) and  (\ref{fcnc down}) can be written in the  form:
\begin{eqnarray}
(\Gamma^u_h)_{ij} & = & \sqrt{2}\frac{m_j}{v_{SM}}\left[\delta_{ij} 
+\alpha_1 (U_L^u)_{i1}(U_L^{u\dagger})_{1j}
-(U_L^u)_{i4}(U_L^{u\dagger})_{4j}\right. \label{fcnc up2}\\
&&+\left. \alpha_2 (U_L^u)_{i1} (U_L^{u\dagger})_{4j}
+\alpha_3 (U_L^u)_{i4} (U_L^{u\dagger})_{1j}\right],
\nonumber
\end{eqnarray}
and,
\begin{eqnarray}
(\Gamma^d_h)_{ij} & = & \sqrt{2}\frac{m_j}{v_{SM}}\left\{\delta_{ij}
+\beta_1 \left[(U_L^u)_{i2}(U_L^{u\dagger})_{2j}+(U_L^u)_{i3}(U_L^{u\dagger})_{3j}\right]
-\left[(U_L^d)_{i4}(U_L^{d\dagger})_{4j}+ (U_L^d)_{i5}(U_L^{d\dagger})_{5j}\right]\right.
\nonumber\\
&&+\left. \beta_2 \left[(U_L^{d})_{i2}(U_L^{d\dagger})_{4j}
+(U_L^{d})_{i3}(U_L^{d\dagger})_{5j}\right]
+\beta_3 \left[(U_L^{d})_{i4}(U_L^{d\dagger})_{2j}
+(U_L^{d})_{i5}(U_L^{d\dagger})_{3j}\right]\right\},
\label{fcnc down2}
\end{eqnarray}
where $\alpha_i$ and $\beta_i$  are $\mathcal{O}(v_{light}/v_{heavy})$.  

One can now see that most of the off-diagonal terms in Eqs. (\ref{fcnc up2}) and (\ref{fcnc down2}) are acquiring strong  suppression from the cofficients $\alpha_i$ and $\beta_i$. Only the third terms in Eqs. (\ref{fcnc up2}) and (\ref{fcnc down2}) are not suppressed by them. Those terms explicitly depend on the left-handed diagonalization matrix elements. Therefore, those elements have to be small in order to make sure the that the FCNCs are suppressed.    

\subsection{Suppression of FCNCs by FN mechanism}

The off-diagonal elements of the quark left-handed  diagonalization matrices are small if the  quark mass matrices satisfy the following hierarchy,
\begin{equation}\label{quark mass matrix hierarchy constraint}
m^q_{i,j}\leq m^q_{i+1,j},
\end{equation}
where $q=u,d$.
In our FN setting this translates into the condition,
\be\label{hierarchy condition for effective charges}
a^q_{i+1}\leq a^q_{i},
\ee
where $a^q_{i}$ are the effective left-handed FN-charges presented in Eqs. (\ref{effective charges up}) and (\ref{effective charges down}).
The left-handed rotation matrices now satisfy,
\begin{equation}\label{UL texture formula}
(U_L^q)_{ij} \sim \epsilon^{\lvert a^q_i-a^q_j\rvert}.
\end{equation}
This may provide additional suppression to the FCNCs. Note that the off-diagonal contributions in Eqs. (\ref{fcnc up2}) and (\ref{fcnc down2}) depend only on the left-handed rotation matrices. The FN-charge assignment of the left-handed SM quarks  determines the amount of suppression (the FN-charges of the left-handed exotic quarks are determined by left-handed SM quarks as they reside in the same representation).

Let us consider the conditions required to satisfy Eq. (\ref{hierarchy condition for effective charges}). The left-handed quark FN charges have to minimally  satisfy:
\be\label{left-handed quark FN charge condition}
(Q^c_{L,3})\leq (Q^c_{L,2})\leq (Q^c_{L,1}),
\ee
to satisfy  Eqs. (\ref{effective charges up}) and (\ref{effective charges down}). 
In addition the scale of the $SU(3)_L\times U(1)_X$-breaking has to be large enough to satisfy, 
\be\label{exotic quarks heavier than SM quarks condition}
a^q_{4}\leq a^q_{3},
\ee 
which ensures that the exotic quarks are not lighter than the SM quarks (also experimentally the exotic quarks have to be heavier than $\mathcal{O}(1\textrm{TeV})$ \cite{Patrignani:2016xqp}).

So the FN charges of the left-handed quarks have to be ordered and the scale of the $SU(3)_L\times U(1)_X$-breaking has to be larger than the electroweak scale.  These are not unreasonable requirements.

 The left-handed charges also determine one important property: the texture of the CKM-matrix. The correct texture for the CKM-matrix can be produced by choosing the value of the  FN expansion parameter $\epsilon$ to be  the Cabibbo angle $\epsilon=0.23$:
\be 
V_{CKM}^{SM}\sim \left(\begin{array}{ccc}
1 & \epsilon^1 & \epsilon^3\\
\epsilon^1 &  1 & \epsilon^2\\
\epsilon^3 & \epsilon^2 & 1
\end{array}
\right).
\ee 
This is quite reasonable assumption to demand from the FN-charges as the CKM-matrix emerges without finetuning with the correct texture underneath. Let us implement this. 
We can produce the correct CKM-texture by choosing the FN-charges of left-handed SM quarks to be:
\be 
(Q^c_{L,1})=3+c,\quad q(Q^c_{L,2})=2+c, \quad q(Q^c_{L,3})=0+c,
\ee
where $c$ is integer number, satisfying Eq. (\ref{left-handed quark FN charge condition}).
Let us now for concreteness  fix the left-handed quark FN charges to be:
 $q(Q^c_{L,1})=2$, $q(Q^c_{L,2})=1$, $q(Q^c_{L,3})=-1$. 
With the given left-handed FN charges the condition in Eq. (\ref{exotic quarks heavier than SM quarks condition})  becomes:
\be
x\equiv\log\left(\frac{v'}{v_2}\right)(\log\epsilon)^{-1}-2 \geq 0,
\ee
which for $v'\sim\mathcal{O}(\textrm{EW})$ gives $v_2\geq \textrm{a few TeV}$. This is the lowest scale one would have picked anyway. 

The left-handed charges   determine the texture of the left-handed quark diagonalization matrix as given in Eq. (\ref{UL texture formula}). The  $U^u_L$ and $U^d_L$  will have the following textures with the chosen left-handed charges:
\begin{equation}\label{left-handed up rotation}
U^u_L\sim \left(\begin{array}{cccc}
1            & \epsilon^{1} & \epsilon^{3} & \epsilon^{3+x}\\
\epsilon^{1} & 1            & \epsilon^{2} & \epsilon^{2+x}\\
\epsilon^{3} & \epsilon^{2} & 1            & \epsilon^{x}\\
\epsilon^{3+x} & \epsilon^{2+x} & \epsilon^{x} & 1
\end{array}\right),
\end{equation}
and
\begin{equation}\label{left-handed down rotation}
U^d_L\sim \left(\begin{array}{ccccc}
1            & \epsilon^{1} & \epsilon^{3} & \epsilon^{3+x} & \epsilon^{5+x}\\
\epsilon^{1} & 1            & \epsilon^{2} & \epsilon^{2+x} & \epsilon^{4+x}\\
\epsilon^{3} & \epsilon^{2} & 1            & \epsilon^{x} & \epsilon^{2+x}\\
\epsilon^{3+x} & \epsilon^{2+x} & \epsilon^{x} & 1        & \epsilon^{2}\\
\epsilon^{5+x} & \epsilon^{4+x} & \epsilon^{2+x}  & \epsilon^{2}   & 1\\
\end{array}\right),
\end{equation}
where the suppression of the off-diagonal elements is clearly visible. The quantity x increases with the  scale of the $SU(3)_L\times U(1)_X$ breaking which leads to more substantial suppression at scales higher than a  few TeV.  

So by demanding  that the CKM-matrix  emerges  naturally, one obtains suppression for the FCNCs of quarks.   The FCNCs get suppressed by doing the usual FN-mechanism for quarks, without artificial fine-tuning.

We can now estimate the textures of the physical Yukawa couplings of Higgs, $\Gamma^u_h$ and $\Gamma^d_h$, by utilizing the quark diagonalization matrix textures in Eqs. (\ref{left-handed up rotation}) and (\ref{left-handed down rotation}):

\begin{equation}\label{up quark interaction texture}
\Gamma^u_h\sim\left(\begin{array}{llll}
y_u & y_c[ \epsilon^1\delta] & y_t[\epsilon^{-2}\delta^2] & \frac{m_U}{v_{heavy}}\\
y_u[\epsilon^1\delta ] & y_c & y_t[\epsilon^{-2}\delta^2] & \frac{m_U}{v_{heavy}}\\
y_u[\epsilon^{-2}\delta^2] & y_c[\epsilon^{-2}\delta^2] & y_t & \frac{m_U}{v_{heavy}}\epsilon^{-2}\\
y_u[\delta ] & y_c [\delta ] & y_t[ \epsilon^{-2}\delta] & \frac{m_U}{v_{heavy}}\epsilon^{-4}\delta
\end{array}\right),
\end{equation}
and,
\begin{equation}\label{down quark interaction texture}
\Gamma^d_h\sim\left(\begin{array}{lllll}
y_d & y_s[\epsilon^1\delta ] & y_b[\epsilon^{-1}\delta^2 ] & \frac{m_{D_1}}{v_{heavy}}\epsilon^1 & \frac{m_{D_2}}{v_{heavy}}\epsilon^3\\
y_d[\epsilon^1\delta ] & y_s & y_b[ \epsilon^{-2}\delta^2] & \frac{m_{D_1}}{v_{heavy}} & \frac{m_{D_2}}{v_{heavy}}\epsilon^2\\
y_d[\epsilon^{-1}\delta^2] & y_s[\epsilon^{-2}\delta^2 ] & y_b & \frac{m_{D_1}}{v_{heavy}}\epsilon^{-2} & \frac{m_{D_2}}{v_{heavy}}\\
y_d[ \epsilon^{1}\delta] & y_s[\delta] & y_b[\epsilon^{-2}\delta] &\frac{m_{D_1}}{v_{heavy}}\delta\epsilon^{-4} & \frac{m_{D_2}}{v_{heavy}}\epsilon^{-2}\delta\\
y_d[\epsilon^{3}\delta ] & y_s[\epsilon^2\delta] & y_b[\delta] & \frac{m_{D_1}}{v_{heavy}}\epsilon^{-2}\delta & \frac{m_{D_2}}{v_{heavy}}\delta
\end{array}\right),
\end{equation}
where $\delta=\mathcal{O}(v_{light}/v_{heavy})$. The exotic quark masses are proportional to $v_{\textrm{heavy}}$.  The Yukawa-couplings depending on the masses of the exotic quarks do not vanish at high values of $SU(3)_L\times U(1)_X$ breaking. The order of magnitude of the exotic quarks is known once also the right-handed quark  FN-charges are fixed.

The quantities in square brackets provide suppression of off-diagonal elements. 
The terms in square brackets should be compared to the bounds in the Tables \ref{quark flavour bounds},  \ref{quark flavour bounds2} and 
\ref{exotic quark flavour bounds}.
 One can see from the Tables \ref{quark flavour bounds},  \ref{quark flavour bounds2} and 
\ref{exotic quark flavour bounds} that the entries under the diagonal pass the flavour bounds automatically, since they are proportional to the lighter 
quark masses. We can place tentative bound on the order of magnitude on the ratio of the $SU(2)_L\times U(1)_Y$- and $SU(3)_L\times U(1)_X$-breaking VEVs, $\delta$,  by 
demanding that the  textures in Eqs. (\ref{up quark interaction texture}) and (\ref{down quark interaction texture}) satisfy the flavour bounds. 
The most stringent bound comes from the $(\Gamma^d_h)_{db}$ -element and gives:
\begin{equation}
\delta\lesssim 0.04.
\end{equation}
Assuming $v_{light}=\mathcal{O}(200\textrm{GeV})$, this translates into 
\begin{equation}
v_{heavy}\gtrsim 5\textrm{ TeV}.
\end{equation}

 In our model there are four up-type quarks and five down-type quarks. This means that the "CKM"-matrix is not a square matrix in this model, but a $4\times 5$-matrix. The $W^+_\mu$-boson coupling to quarks is given by, 
 \begin{equation}
\mathcal{L}_{gCC}=\frac{g_{3}}{\sqrt{2}}\bar{u}_L\gamma^{\mu}V_{CKM}^{331}d_L {W}^{+}_{\mu}+h.c.,
\end{equation}
where $V_{CKM}^{331}$ is a $4\times 5$-matrix. Details of the charged currents are given in the Appendix \ref{charged currents}. The CKM-matrix texture with the chosen left-handed quark FN-charges is:
\begin{equation}
V_{ckm}^{331}\sim \left(\begin{array}{ccccc}
1            & \epsilon^{1} & \epsilon^{3}        & \epsilon^{1}\delta    & \epsilon^{3}\delta\\
\epsilon^{1} & 1            & \epsilon^{2}        & \delta                & \epsilon^{2}\delta\\
\epsilon^{3} & \epsilon^{2} & 1                   & \epsilon^{-2}\delta   & \delta\\
\delta       & \delta       & \epsilon^{-1}\delta & \epsilon^{-4}\delta^2 & \epsilon^{-2}\delta^2\\
\end{array}\right).
\end{equation}
The $3\times 3$-block in the upper-left corner corresponds to the CKM-matrix of the Standard Model. The $W_\mu^\pm$ boson  couples to exotic quarks. The  $W_\mu^\pm$  boson, therefore, has additional flavour violating contributions to the neutral meson mixing. We find the $W_\mu^\pm$ mediated BSM contribution to the neutral meson mixing to be always subleading to Higgs-mediated. Next we will consider Higgs mediated neutral meson mixing which provides the  tightest constraints for the flavour violating couplings.
We have also checked other quark flavour violating processes like leptonic decay of neutral meson, $M^0\to l_i^+ l_i^-$, radiative B-meson decay, $\bar{B}^0\to X_s^0\gamma$, and top quark decays, $t\to h c$ and $t\to q\gamma$. Top quark decay is considered in the appendix \ref{top quark decay appendix}, as it gives direct access to the top Yukawa coupling at tree-level. The bound is rather weak, however. The leptonic meson decay and   B-meson decay provide bounds that are weaker than the neutral meson mixing in our examples.

\section{Higgs mediated neutral meson mixing}\label{Higgs mediated neutral meson mixing}
The flavour changing neutral currents manifest themselves at quark sector in 331-models.  Here we list the most important processes  that pose constraints to Higgs-quark Yukawa couplings. Here we assume that the Higgs is the only significant  source of  flavour violation beyond the Standard Model.
The neutral mesons\footnote{The neutral mesons and their quark content: $K^0=d\bar{s}$, $B_d^0=d\bar{b}$,$B_s^0=s\bar{b}$, $D^0=c\bar{u}$} can mix with their antiparticles in the Standard Model due to flavour violation in the $W^{\pm}_\mu$ coupling to quarks. The BSM contribution to neutral meson mixing is tightly constrained by the experiments.
The Higgs mediates neutral meson mixing at tree-level which has been a  problem  in the usual 331-models as they provide no suppression mechanism. Our FN331-model that incorporates the FN-mechanism into the 331-setting can quite naturally  provide the neccesary suppression as we have already demonstrated.  
  
The most general effective Hamiltonians describing neutral meson mixing have the following form:
\begin{eqnarray}
\mathcal{H}^{\Delta S=2}_{eff} & = & \sum_{k=1}^5 C_k^{ds} Q^{ds}_k + \sum_{k=1}^3 \widetilde{C}_k^{ds} \widetilde{Q}^{ds}_k,\nonumber\\
\mathcal{H}^{\Delta C=2}_{eff} & = & \sum_{k=1}^5 C_k^{cu} Q^{cu}_k + \sum_{k=1}^3 \widetilde{C}_k^{cu} \widetilde{Q}^{cu}_k,\nonumber\\
\mathcal{H}^{\Delta B=2}_{eff} & = & \sum_{q=d,s}\sum_{k=1}^5 C_k^{qb} Q^{qb}_k + \sum_{q=d,s}\sum_{k=1}^3 \widetilde{C}_k^{qb} \widetilde{Q}^{qb}_i,\nonumber
\end{eqnarray}
where $C_k^{qq'}$s are Wilson coefficients and,
\begin{eqnarray}
Q^{q_i q_j}_1 & = & (\bar{q}_{j,L}^\alpha\gamma^{\mu}q^\alpha_{i,L})(\bar{q}_{j,L}^\beta\gamma_{\mu}q^\beta_{i,L}),\nonumber\\ 
Q^{q_i q_j}_2 & = & (\bar{q}_{j,R}^\alpha q^\alpha_{i,L})(\bar{q}_{j,R}^\beta q^\beta_{i,L}),\nonumber\\ 
Q^{q_i q_j}_3 & = & (\bar{q}_{j,R}^\alpha q^\beta_{i,L})(\bar{q}_{j,R}^\beta q^\alpha_{i,L}),\nonumber\\ 
Q^{q_i q_j}_4 & = & (\bar{q}_{j,R}^\alpha q^\alpha_{i,L})(\bar{q}_{j,L}^\beta q^\beta_{i,R}),\nonumber\\ 
Q^{q_i q_j}_5 & = & (\bar{q}_{j,R}^\alpha q^\beta_{i,L})(\bar{q}_{j,L}^\beta q^\alpha_{i,R}),\nonumber
\end{eqnarray}
are the four-fermion operators.
The $\alpha$ and $\beta$ are colour indices. The operators $\widetilde{Q}^{q_i q_j}_{1,2,3}$ are obtained from the operators $Q^{q_i q_j}_{1,2,3}$ by replacement $L\leftrightarrow R$. 

One can only bound   the Yukawa couplings involving the light SM quarks, u,d,s,c and b,  by tree-level processes. The top quark and the exotic quarks $U$, $D_1$ and $D_2$ enter the neutral meson mixing  processes at one-loop level. 
One has   to turn  1-loop processes in order to acquire bounds for Yukawa couplings involving  exotic quarks or the top quarks and even then the bound is obtaned for a product   of multiple couplings. The bounds on those Yukawa couplings are thus very loose. We provide the details of the 1-loop contributions to neutral meson mixing in the appendix \ref{neutral meson mixing at 1 loop}.  

The processes $K^0$-$\bar{K}^0$, $B_d^0$-$\bar{B_d}^0$ and $B_s^0$-$\bar{B_s}^0$ provide bounds on all the off-diagonal components of $\Gamma^d_h$ involving Standard Model quarks at tree-level.
The $D^0$-$\bar{D}^0$ gives bounds on elements $(\Gamma^u_h)_{uc}$  and $(\Gamma^u_h)_{cu}$ in the up-sector at tree-level. The top quark does not hadronize, and there is no bound on off-diagonal  $(\Gamma^u_h)$-elements involving top from tree-level.
 
\subsection{Tree-level}
The effective Hamiltonians describing the neutral meson  mixing at tree-level are:
\begin{equation}
\mathcal{H}_{eff}(K^0\textrm{-}\bar{K}^0)=C^{ds}_2 (\bar{s}_R d_L)^2 + \widetilde{C}^{ds}_2 (\bar{s}_L d_R)^2 
+ C^{ds}_4 (\bar{s}_R d_L)(\bar{s}_L d_R),
\end{equation}
\begin{equation}
\mathcal{H}_{eff}(B_d^0\textrm{-}\bar{B}_d^0)=C^{db}_2 (\bar{b}_R d_L)^2 + \widetilde{C}^{db}_2 (\bar{b}_L d_R)^2 
+ C^{db}_4 (\bar{b}_R d_L)(\bar{b}_L d_R),
\end{equation}
\begin{equation}
\mathcal{H}_{eff}(B_s^0\textrm{-}\bar{B}_s^0)=C^{sb}_2 (\bar{b}_R s_L)^2 + \widetilde{C}^{sb}_2 (\bar{b}_L s_R)^2 
+ C^{sb}_4 (\bar{b}_R s_L)(\bar{b}_L s_R),
\end{equation}
\begin{equation}
\mathcal{H}_{eff}(D^0\textrm{-}\bar{D}^0)=C^{cu}_2 (\bar{u}_R c_L)^2 + \widetilde{C}^{uc}_2 (\bar{u}_L c_R)^2 
+ C^{cu}_4 (\bar{u}_R c_L)(\bar{u}_L c_R).
\end{equation}
The operators  $Q_1$, $Q_3$ and $Q_5$ are not generated at tree-level.
The diagrams contributing to neutral meson mixing at tree-level are given in the Figure \ref{neutral meson mixing}. The Wilson coefficients for neutral kaon mixing are:
\begin{eqnarray}
C^{ds}_2=-\frac{(\Gamma^{d\ast}_h)^2_{ds}}{4m^2_h},\quad \widetilde{C}^{ds}_2=-\frac{(\Gamma^{d}_h)^2_{sd}}{4m^2_h},\quad
C^{ds}_4=-\frac{(\Gamma^{d\ast}_h)_{ds}(\Gamma^d_h)_{sd}}{2m^2_h}.\label{kaon mixing}
\end{eqnarray}
The Wilson coefficients for  other processes are obtained from Eq. (\ref{kaon mixing}), by  replacing quark flavour indices. The experimental constraints on the Wilson coefficients can be found in \cite{Bona:2007vi}. The current experimental bounds on flavour changing quark Yukawa couplings are given in the 
Tables  \ref{quark flavour bounds}, \ref{quark flavour bounds2} and \ref{exotic quark flavour bounds}. 

\begin{minipage}{\linewidth}
\begin{figure}[H]
\begin{tikzpicture}[thick,scale=1.0]
\draw[particle] (-1,1) -- node[black,above,xshift=-0.6cm,yshift=0.4cm] {$d_{L}$} (0,0);
\draw[particle] (0,0) -- node[black,above,xshift=-0.5cm,yshift=-1.0cm] {$s_{R}$} (-1,-1);
\draw[particle] (1,0) -- node[black,above,yshift=0.4cm,xshift=0.6cm] {$s_{R}$} (2,1);
\draw[particle] (2,-1) -- node[black,above,yshift=-1.0cm,xshift=0.6cm] {$d_{L}$} (1,0);
\draw[dashed] (0,0) -- node[black,above,yshift=0.0cm,xshift=-0.0cm] {$h$} (1,0);
\end{tikzpicture}
\hspace{5mm}
\begin{tikzpicture}[thick,scale=1.0]
\draw[particle] (-1,1) -- node[black,above,xshift=-0.6cm,yshift=0.4cm] {$d_{R}$} (0,0);
\draw[particle] (0,0) -- node[black,above,xshift=-0.5cm,yshift=-1.0cm] {$s_{L}$} (-1,-1);
\draw[particle] (1,0) -- node[black,above,yshift=0.4cm,xshift=0.6cm] {$s_{L}$} (2,1);
\draw[particle] (2,-1) -- node[black,above,yshift=-1.0cm,xshift=0.6cm] {$d_{R}$} (1,0);
\draw[dashed] (0,0) -- node[black,above,yshift=0.0cm,xshift=-0.0cm] {$h$} (1,0);
\end{tikzpicture}
\hspace{5mm}
\begin{tikzpicture}[thick,scale=1.0]
\draw[particle] (-1,1) -- node[black,above,xshift=-0.6cm,yshift=0.4cm] {$d_{L}$} (0,0);
\draw[particle] (0,0) -- node[black,above,xshift=-0.5cm,yshift=-1.0cm] {$s_{R}$} (-1,-1);
\draw[particle] (1,0) -- node[black,above,yshift=0.4cm,xshift=0.6cm] {$s_{L}$} (2,1);
\draw[particle] (2,-1) -- node[black,above,yshift=-1.0cm,xshift=0.6cm] {$d_{R}$} (1,0);
\draw[dashed] (0,0) -- node[black,above,yshift=0.0cm,xshift=-0.0cm] {$h$} (1,0);
\end{tikzpicture}
\vspace{0.5cm}
\caption{The diagrams contributing to $K^0$-$\bar{K}^0$ mixing at tree-level. The corresponding diagrams  for other neutral mesons are obtained by obvious quark flavour replacements.}
\label{neutral meson mixing} 
\end{figure}
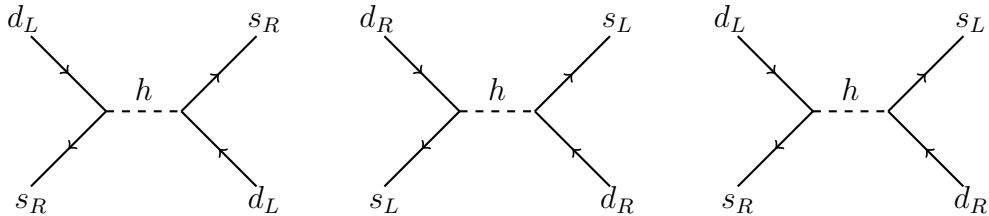    
\end{minipage}

 \vspace{0.3cm}

\begin{table}[h!]
\begin{center}
 {\tabulinesep=1.2mm
   \begin{tabu} {c|c|c}
       Process  & coupling & Current bound\\
\hline\hline
$K^0-\bar{K}^0$ (tree-level)& $ \sqrt{\textrm{Re}({\Gamma^d_{ds}}^2)}$  &  $[-3.45,3.35]\times 10^{-5}=\big([-6.3,6.1]\times 10^{-2}\big)\times y_s$
\\[2mm]
                                 & $ \sqrt{\textrm{Re}({\Gamma^d_{sd}}^2)}$  & $[-3.45,3.35]\times 10^{-5}=[-1.25,1.22]\times y_d$
\\[2mm]
                     & $ \sqrt{\textrm{Re}({\Gamma^d_{sd}}^{\ast}\Gamma^d_{ds})}$  & $[-1.1,1.1]\times 10^{-5}=\big([-8.6,8.6]\times 10^{-2}\big)\times \sqrt{y_d y_s}$
\\[2mm]
\hline
$B^0_d-\bar{B}^0_d$ (tree-level) & $\lvert \Gamma^d_{db}\rvert$  & $<2.12\times 10^{-4}=\big(8.8\times 10^{-3}\big)\times y_b$
\\[2mm]
                               & $\lvert \Gamma^d_{bd}\rvert$  & $<2.12\times 10^{-4}=7.7\times y_d$
\\[2mm]
                     & $ \sqrt{\lvert{\Gamma^d_{sd}}^{}\Gamma^d_{ds}\rvert}$  & $<8.1\times 10^{-5}=\big(1.0\times 10^{-1})\times \sqrt{y_d y_b}$
\\[2mm]
\hline
$B_s^0-\bar{B}_s^0$ (tree-level) & $\lvert \Gamma^d_{sb}\rvert$  & $<1.9\times10^{-3}=\big(7.8\times 10^{-2}\big)\times y_b$ 
\\[2mm]
                                      & $\lvert \Gamma^d_{bs}\rvert$  & $<1.9\times10^{-3}=3.4\times y_s$ 
\\[2mm]
                     & $ \sqrt{\lvert{\Gamma^d_{sd}}^{}\Gamma^d_{ds}\rvert}$  & $<7.1\times 10^{-4}=\big(2.0\times 10^{-1})\times \sqrt{y_s y_b}$
\\[2mm]
   \end{tabu}}
\caption{Current experimental bounds on flavour changing Yukawa couplings of quarks \cite{Bona:2007vi}.}
\label{quark flavour bounds}
\end{center}
\end{table}

\newpage
\begin{table}[h!]
\begin{center}
 {\tabulinesep=1.2mm
   \begin{tabu} {c|c|c}
       Process  & coupling & Current bound\\
\hline\hline
 $D^0-\bar{D}^0$ \textrm{(Tree-level)}  & $\lvert \Gamma^u_{uc}\rvert$ & $<1.0\times 10^{-4}=\big(1.4\times 10^{-2}\big)\times y_c$ 
\\[2mm]
                                                   &  $\lvert \Gamma^u_{cu}\rvert$ & $<1.0\times 10^{-4}=7.6\times y_u$ 
\\[2mm]
                     & $ \sqrt{\lvert{\Gamma^u_{cu}}^{}\Gamma^u_{uc}\rvert}$  & $<3.9\times 10^{-5}=\big(1.2\times 10^{-1})\times \sqrt{y_u y_c}$
\\[2mm]
\hline
$t\to hc$              &  $\sqrt{\lvert \Gamma^u_{ct}\rvert^2+\lvert \Gamma^u_{tc}\rvert^2}$   &     $<0.24=(0.24)\times  y_t$\\[2mm]
\hline
                                       & $\lvert \Gamma^u_{ut}\Gamma^u_{ct}\rvert$ & $<1.6\times 10^{-2}=\big(1.6\times 10^{-2}\big)\times y_t^2$ 
\\[2mm]

 $D^0-\bar{D}^0$ \textrm{(Loop-level)}  &   $\lvert \Gamma^u_{ut}\Gamma^u_{tc}\rvert$ & $<4.4\times 10^{-3}=\big(6.1\times 10^{-1}\big) \times y_c y_t$ 
\\[2mm]
                                                  &  $\lvert \Gamma^u_{ut}\Gamma^u_{tu}\Gamma^u_{ct}\Gamma^u_{tc}\rvert^{1/2}$ & $<1.7\times 10^{-3}=5.5\times y_t\sqrt{y_u y_c}$ 
\\[2mm]
   \end{tabu}}
\end{center}
\caption{Current experimental bounds on flavour changing Yukawa couplings involving exotic quarks $U$, $D_1$ and $D_2$ \cite{Bona:2007vi}.~~~~~~~~~~~~~~~~~~~~~~~~~~~~~~~~~~~~~~~~~~~~~~~~~~~~~~~~~~~~~~~~~~~~~~~~~~~~~~~~~}
\label{quark flavour bounds2}
\end{table}

\begin{table}[h!]
\begin{center}
 {\tabulinesep=1.2mm
   \begin{tabu} {c|c|c}
       Process  & coupling & Current bound\\
\hline\hline
$K^0-\bar{K}^0$ (Loop-level) & $ \sqrt{\textrm{Re}[(\Gamma^d_{sD}\Gamma^{d\ast}_{dD})^2]}\frac{\textrm{TeV}}{m_D}$  &  $[-7.0,7.0]\times 10^{-2}$
\\[2mm]
\hline
$B^0-\bar{B}^0$ (Loop-level) & $\lvert \Gamma^d_{bD}\Gamma^d_{dD}\rvert \frac{\textrm{TeV}}{m_D}$  & $<3.4\times 10^{-1}$
\\[2mm]
\hline
$B_s^0-\bar{B}_s^0$ (Loop-level) & $\lvert \Gamma^d_{bD}\Gamma^d_{sD}\rvert \frac{\textrm{TeV}}{m_D}$  & $<2.4$
\\[2mm]
\hline
$D^0-\bar{D}^0$ (Loop-level) & $\lvert \Gamma^u_{uU}\Gamma^u_{cU}\rvert \frac{\textrm{TeV}}{m_U}$  & $<6.0\times 10^{-2}$\\[2mm]
   \end{tabu}}
\caption{Current experimental bounds on flavour changing Yukawa couplings involving exotic quarks $U$, $D_1$ and $D_2$ \cite{Bona:2007vi}. Here we have assumed $\Gamma_{ij}=(m_j/m_i)\Gamma_{ji}$, as suggested by the textures in Eqs. (\ref{fcnc up}) and (\ref{fcnc down}).~~~~~~~~~~~~~~~~~~~~~~~~~~~~~~~~~~~~~~~~~~~~~~~~~~~~~~~~~~~~~~~~~~~~~~~~~~~~~~~~~}
\label{exotic quark flavour bounds}
\end{center}
\end{table}

We will now present numerical examples with different FN charges for quarks and demonstrate the supression of FCNCs.

\section{Numerical examples}\label{numerical examples}
In the numerical examples we generate up- and down-type quark mass matrices by demanding the correct masses for the SM quarks. Furthermore, we demand that the magnitude of the SM part of  CKM-matrix elements match the experimental values at $2\sigma$ confidence level. All the Yukawa matrix elements are in principle complex numbers and therefore induce CP-violating effects. We have chosen all our Yukawa matrices to have real entries for simplicity of the numerical analysis, as we are only interested in the flavour violation.    
We present three different examples that differ by the up- and down-type quark textures and  the vacuum expectation values. The ballpark of the $SU(3)_L\times U(1)_X$ breaking VEVs are  in the examples chosen so that it is as small as possible while  still producing the  lightest exotic quark  mass over 1 TeV to avoid limits  exotic quark searches impose \cite{Patrignani:2016xqp}. The $SU(3)_L\times U(1)_X$ breaking scale can thus change depending on the details of the FN charges. 

We will consider three qualitatively different examples for the quark mass matrix textures. 
In all of the three examples we will fix the left-handed quark FN charges to be:
 $q(Q^c_{L,1})=2$, $q(Q^c_{L,2})=1$, $q(Q^c_{L,3})=-1$. This charge assignment will produce the correct texture for the CKM-matrix.
The FN charges of the right-handed quarks differ in each example. 

\subsection{Example 1}
The FN charges of the right-handed quarks are: $q(u_{R,1})=5$,  $q(u_{R,2})=2$,  $q(u_{R,3})=0$, $q(U_{R})=0$,  $q(d_{R,1})=7$,  $q(d_{R,2})=5$, $q(d_{R,3})=4$, $q(D_{R,1})=3$ and $q(D_{R,2})=2$.
With these FN-charges the quark mass matrix textures are,
\begin{equation}
m_u\sim \left(\begin{array}{cccc}
v_{light}~\epsilon^8        & v_{light}~\epsilon^5 & v_{light}~\epsilon^3 & v_{light}~\epsilon^3\\
v_{light}~\epsilon^7 & v_{light}~\epsilon^4            & v_{light}~\epsilon^2 & v_{light}~\epsilon^2\\
v_{light}~\epsilon^5 & v_{light}~\epsilon^2 & v_{light}~\epsilon^0            & v_{light}~\epsilon^0\\
v_{heavy}~\epsilon^7 & v_{heavy}~\epsilon^4 & v_{heavy} ~\epsilon^2 & v_{heavy}~\epsilon^2
\end{array}\right),
\end{equation}
and
\begin{equation}
m_d\sim \left(\begin{array}{ccccc}
v_{light}~\epsilon^8        & v_{light}~\epsilon^6 & v_{light}~\epsilon^5 & v_{light}~\epsilon^4 & v_{light}~\epsilon^3 \\
v_{light}~\epsilon^7        & v_{light}~\epsilon^5 & v_{light}~\epsilon^4 & v_{light}~\epsilon^3 & v_{light}~\epsilon^2 \\
v_{light}~\epsilon^5        & v_{light}~\epsilon^3 & v_{light}~\epsilon^2 & v_{light}~\epsilon^1 & v_{light}~\epsilon^0 \\
v_{heavy}~\epsilon^7 & v_{heavy}~\epsilon^5 & v_{heavy} ~\epsilon^4 & v_{heavy}~\epsilon^3  & v_{heavy}~\epsilon^2 \\
v_{heavy}~\epsilon^5 & v_{heavy}~\epsilon^3 & v_{heavy} ~\epsilon^2 & v_{heavy}~\epsilon^1  & v_{heavy}~\epsilon^0 \\
\end{array}\right).
\end{equation}
The exotic quark masses schematically are: 
\be
m_U\sim\epsilon^2 v_{\rm heavy}, \quad m_{D_1}\sim\epsilon^3 v_{\rm heavy}, \quad m_{D_2}\sim\epsilon^0 v_{\rm heavy}.
\ee
The lightest exotic quark mass is suppressed by $\epsilon^3$ compared to $v_{\rm heavy}$.  In order to have $m_{D_1}$ around 1 TeV the $v_{\rm heavy}$ has to  be $v_{\rm heavy}\sim 50$ TeV. The rest of the exotic quarks are less suppressed and are heavier.  

The numerical values for the $SU(3)_L\times U(1)_X$-breaking VEVs are $u=48000$ GeV and $v_2=55000$ GeV, and the $SU(2)_L\times U(1)_Y$-breaking VEVs are $v_1=100$ GeV and $v'=237.05$ GeV. The used order-one coefficients and scalar potential parameters are presented in the Appendix \ref{numerical example 1}. The exotic quark masses are: $m_U=5$ TeV, $m_{D_1}=1.303$ TeV and $m_{D_2}=50.7$ TeV. The physical Higgs Yukawa couplings are:
\begin{equation}\label{physical up yukawa numerical 1}
\Gamma^u_h=\left(
\begin{array}{cccc}
 1.1\times 10^{-5} & 1.1\times 10^{-9} & -5.0\times 10^{-5} & -2.6\times 10^{-2} \\
 2.2\times 10^{-11} & 7.3\times 10^{-3} & -5.8\times 10^{-5} & -2.9\times 10^{-2} \\
 -5.9\times 10^{-10} & 3.9\times 10^{-7} & -9.9\times 10^{-1} & 1.6\\
 1.0\times 10^{-8} & -6.9\times 10^{-6} & -5.7\times 10^{-2}  & 9.3\times 10^{-2} \\
\end{array}
\right),
\end{equation}
and 
\begin{equation}\label{physical down yukawa numerical 1}
\Gamma^d_h=\left(
\begin{array}{ccccc}
 2.7\times 10^{-5} & 2.2\times 10^{-9} & 2.0\times 10^{-8} & -3.8\times 10^{-4} & 8.1\times 10^{-3}\\
 1.1\times 10^{-10} & 5.7\times 10^{-4} & -3.0\times 10^{-7} & 5.2\times 10^{-3} & -2.6\times 10^{-2} \\
 1.8\times 10^{-10} & 1.1\times 10^{-8} & 2.4\times 10^{-2} & 1.3\times 10^{-1} & 3.5\times 10^{-2}\\
 -1.0\times 10^{-8} & -6.2\times 10^{-7} & 4.2\times 10^{-4} & 2.1\times 10^{-3} & 6.3\times 10^{-4} \\
 6.2\times 10^{-11} & -1.6\times 10^{-7} & -4.1\times 10^{-5} & -2.2\times 10^{-4} & -5.2\times 10^{-3} \\
\end{array}
\right).
\end{equation}
  
 The CKM-matrix at $2\sigma$ confidence level is \cite{Patrignani:2016xqp}: 

\begin{equation}\label{CKM matrix numerical 1}
 \lvert V_{ckm}^{331}\rvert=
\left(
\begin{array}{ccccc}
 0.974 & 0.226 & 0.00336 & 0.00015 & 0.000048 \\
 0.23 & 0.97 & 0.0433 & 0.000082 & 0.000087 \\
 0.007 & 0.0429 & 0.997 & 0.017 & 0.00012\\
 0.00072 & 0.0017 & 0.057 & 0.00099 & 6.6\times 10^{-6} \\
\end{array}
\right).
\end{equation}

\subsection{Example 2}
The FN charges of the right-handed quarks are: $q(u_{R,1})=5$,  $q(u_{R,2})=2$,  $q(u_{R,3})=0$, $q(U_{R})=0$,  $q(d_{R,1})=7$,  $q(d_{R,2})=5$, $q(d_{R,3})=4$, $q(D_{R,1})=2$ and $q(D_{R,2})=2$.
With these FN-charges the quark mass matrix textures are,
\begin{equation}
m_u\sim \left(\begin{array}{cccc}
v_{light}~\epsilon^8        & v_{light}~\epsilon^5 & v_{light}~\epsilon^3 & v_{light}~\epsilon^3\\
v_{light}~\epsilon^7 & v_{light}~\epsilon^4            & v_{light}~\epsilon^2 & v_{light}~\epsilon^2\\
v_{light}~\epsilon^5 & v_{light}~\epsilon^2 & v_{light}~\epsilon^0            & v_{light}~\epsilon^0\\
v_{heavy}~\epsilon^7 & v_{heavy}~\epsilon^4 & v_{heavy} ~\epsilon^2 & v_{heavy}~\epsilon^2
\end{array}\right),
\end{equation}
and
\begin{equation}
m_d\sim \left(\begin{array}{ccccc}
v_{light}~\epsilon^8        & v_{light}~\epsilon^6 & v_{light}~\epsilon^5 & v_{light}~\epsilon^3 & v_{light}~\epsilon^3 \\
v_{light}~\epsilon^7        & v_{light}~\epsilon^5 & v_{light}~\epsilon^4 & v_{light}~\epsilon^2 & v_{light}~\epsilon^2 \\
v_{light}~\epsilon^5        & v_{light}~\epsilon^3 & v_{light}~\epsilon^2 & v_{light}~\epsilon^0 & v_{light}~\epsilon^0 \\
v_{heavy}~\epsilon^7 & v_{heavy}~\epsilon^5 & v_{heavy} ~\epsilon^4 & v_{heavy}~\epsilon^2  & v_{heavy}~\epsilon^2 \\
v_{heavy}~\epsilon^5 & v_{heavy}~\epsilon^3 & v_{heavy} ~\epsilon^2 & v_{heavy}~\epsilon^0  & v_{heavy}~\epsilon^0 \\
\end{array}\right).
\end{equation}
The exotic quark masses schematically are: 
\be
m_U\sim\epsilon^2 v_{\rm heavy}, \quad m_{D_1}\sim\epsilon^2 v_{\rm heavy}, \quad m_{D_2}\sim\epsilon^0 v_{\rm heavy}.
\ee
The lightest exotic quark mass is suppressed by $\epsilon^2$ compared to $v_{\rm heavy}$. The $v_{\rm heavy}$ has to thus be $v_{\rm heavy}\sim 20$ TeV in order to have $m_{D_1}$ around 1 TeV. The heaviest exotic quark is not suppressed by $\epsilon$  and will be heavier.

The numerical values for the $SU(3)_L\times U(1)_X$-breaking VEVs are $u=21000$ GeV and $v_2=19000$ GeV, and the $SU(2)_L\times U(1)_Y$-breaking VEVs are $v_1=198$ GeV and 
$v'=198$ GeV. The used order-one coefficients and scalar potential parameters are presented in the Appendix \ref{numerical example 2}. The exotic quark masses are: $m_U=2$ TeV, $m_{D_1}=2.1$ TeV and $m_{D_2}=25.2$ TeV. The physical Higgs Yukawa couplings are: 
\begin{equation}\label{physical up yukawa numerical 2}
\Gamma^u_h=\left(
\begin{array}{cccc}
-1.3\times 10^{-5} & -4.4\times 10^{-7} & 4.7\times 10^{-4} & -4.6\times 10^{-2} \\
 1.8\times 10^{-9} & 7.3\times 10^{-3} & 1.4\times 10^{-3} & -1.4\times 10^{-1} \\
 -1.5\times 10^{-8} & 9.5\times 10^{-6} & 9.8\times 10^{-1} & 1.3\\
 1.3\times 10^{-7} & -8.1\times 10^{-5} & 1.1\times 10^{-1}  & 1.6\times 10^{-1} \\
\end{array}
\right),
\end{equation}
and
\begin{equation}\label{physical down yukawa numerical 2}
\Gamma^d_h=\left(
\begin{array}{ccccc}
2.7\times 10^{-5} & 2.7\times 10^{-9} & 2.3\times 10^{-6} & -1.7\times 10^{-2} & 5.8\times 10^{-2} \\
 -4.1\times 10^{-11} & 5.5\times 10^{-4} & -6.2\times 10^{-6} & 4.4\times 10^{-2} & -1.8\times 10^{-1} \\
 3.6\times 10^{-9} & -3.7\times 10^{-7} & 2.4\times 10^{-2}& 8.5\times 10^{-1} & -1.1 \\
 -5.2\times 10^{-8} & 5.3\times 10^{-6} & 1.7\times 10^{-3} & 5.8\times 10^{-2} & -7.8\times 10^{-2} \\
 6.0\times 10^{-9} & -3.8\times 10^{-7} & -4.0\times 10^{-5} & -1.4\times 10^{-3} & -7.1\times 10^{-3}\\
\end{array}
\right).
\end{equation}
  
 The CKM-matrix at $2\sigma$ confidence level is 

\begin{equation}\label{CKM matrix numerical 2}
V_{ckm}^{331}=
\left(
\begin{array}{ccccc}
-0.974 & -0.225 & 0.00334 & 0.00079 & -0.00014 \\
 0.22 & -0.97 & 0.0434 & -0.00085 & 0.00094 \\
 -0.007 & 0.0411 & 0.999 & 0.069 & -0.0076 \\
 0.00052 & 0.017 & 0.11 & 0.0081& -0.00089 \\
\end{array}
\right).
\end{equation}

\subsection{Example 3}
The FN charges of the right-handed quarks are: $q(u_{R,1})=5$,  $q(u_{R,2})=2$,  $q(u_{R,3})=0$, $q(U_{R})=-2$,  $q(d_{R,1})=7$,  $q(d_{R,2})=5$, $q(d_{R,3})=4$, $q(D_{R,1})=0$ and $q(D_{R,2})=1$.
With these FN-charges the quark mass matrix textures are,
\begin{equation}
m_u\sim \left(\begin{array}{cccc}
v_{light}~\epsilon^8        & v_{light}~\epsilon^5 & v_{light}~\epsilon^3 & v_{light}~\epsilon^1\\
v_{light}~\epsilon^7 & v_{light}~\epsilon^4            & v_{light}~\epsilon^2 & v_{light}~\epsilon^0\\
v_{light}~\epsilon^5 & v_{light}~\epsilon^2 & v_{light}~\epsilon^0            & v_{light}~\epsilon^2\\
v_{heavy}~\epsilon^7 & v_{heavy}~\epsilon^4 & v_{heavy} ~\epsilon^2 & v_{heavy}~\epsilon^0
\end{array}\right),
\end{equation}
and
\begin{equation}
m_d\sim \left(\begin{array}{ccccc}
v_{light}~\epsilon^8        & v_{light}~\epsilon^6 & v_{light}~\epsilon^5 & v_{light}~\epsilon^1 & v_{light}~\epsilon^2 \\
v_{light}~\epsilon^7        & v_{light}~\epsilon^5 & v_{light}~\epsilon^4 & v_{light}~\epsilon^0 & v_{light}~\epsilon^1 \\
v_{light}~\epsilon^5        & v_{light}~\epsilon^3 & v_{light}~\epsilon^2 & v_{light}~\epsilon^2 & v_{light}~\epsilon^1 \\
v_{heavy}~\epsilon^7 & v_{heavy}~\epsilon^5 & v_{heavy} ~\epsilon^4 & v_{heavy}~\epsilon^0  & v_{heavy}~\epsilon^1 \\
v_{heavy}~\epsilon^5 & v_{heavy}~\epsilon^3 & v_{heavy} ~\epsilon^2 & v_{heavy}~\epsilon^1  & v_{heavy}~\epsilon^0 \\
\end{array}\right).
\end{equation}
The exotic quark masses schematically are: 
\be
m_U\sim\epsilon^0 v_{\rm heavy}, \quad m_{D_1}\sim\epsilon^0 v_{\rm heavy}, \quad m_{D_2}\sim\epsilon^0 v_{\rm heavy}.
\ee
The exotic quark masses are not suppressed by powers of $\epsilon$ so the $v_{\rm heavy}$ can be low. 

The numerical values for the $SU(3)_L\times U(1)_X$-breaking VEVs are $u=5000$ GeV and $v_2=5500$ GeV, and the $SU(2)_L\times U(1)_Y$-breaking VEVs are $v_1=204$ GeV and $v'=204$ GeV. The used order-one coefficients and scalar potential parameters are presented in the Appendix \ref{numerical example 3}. The exotic quark masses are: $m_U=5$ TeV, $m_{D_1}=7$ TeV and $m_{D_2}=8.5$ TeV. The physical Higgs Yukawa couplings are:
\begin{equation}\label{physical up yukawa numerical 3}
\Gamma^u_h=\left(
\begin{array}{cccc}
 1.6\times 10^{-5} & -5.5\times 10^{-6} & -9.8\times 10^{-5} & -5.3\times 10^{-1} \\
 -2.2\times 10^{-8} & 7.3\times 10^{-3} & -2.3\times 10^{-4} & -1.2 \\
 -2.7\times 10^{-9} & -1.5\times 10^{-6} & 9.9\times 10^{-1} & -1.5\times 10^{-1}\\
 -5.1\times 10^{-7} & -2.7\times 10^{-4} & -5.6\times 10^{-3}  & 1.0\times 10^{-1} \\
\end{array}
\right),
\end{equation}
and
\begin{equation}\label{physical down yukawa numerical 3}
\Gamma^d_h=\left(
\begin{array}{ccccc}
2.7\times 10^{-5} & -9.4\times 10^{-9} & -2.5\times 10^{-7} & 8.4\times 10^{-2} & 9.1\times 10^{-2}\\
-1.6\times 10^{-9} & 5.7\times 10^{-4} & -2.0\times 10^{-6} & 5.5\times 10^{-1} & 7.0\times 10^{-1}\\
 3.2\times 10^{-10} & 1.4\times 10^{-8} & -2.4\times 10^{-2} & -5.2\times 10^{-1} & 3.4\times 10^{-1} \\
 6.4\times 10^{-8} & 2.2\times 10^{-6} & 5.7\times 10^{-4} & 7.3\times 10^{-2} & -5.0\times 10^{-3}\\
 4.5\times 10^{-8} & 2.2\times 10^{-6} & -4.1\times 10^{-4} & -6.5\times 10^{-3} & 8.0\times 10^{-2} \\
\end{array}
\right).
\end{equation}
  
 The CKM-matrix at $2\sigma$ confidence level is 

\begin{equation}\label{CKM matrix numerical 3}
V_{ckm}^{331}=
\left(
\begin{array}{ccccc}
 -0.974 & -0.226 & 0.00341 & -0.0052 & -0.0051 \\
 0.23 & -0.97 & 0.0435 & -0.014 & -0.013 \\
 0.0066 & -0.043 & -0.999 & 0.012 & -0.0075 \\
 0.0088 & 0.043 & 0.0036 & 0.00057 & 0.00067 \\
\end{array}
\right).
\end{equation}

All the numerical examples we have presented satisfy the flavour bounds.

\section{Conclusion}
\label{conclusion}
 The model discussed here introduces the Froggatt-Nielsen mechanism into a 331-model with $\beta=\pm 1/\sqrt{3}$ in  such a way that no  scalars are required other than those already present in the traditional 331-models with $\beta=\pm 1/\sqrt{3}$. The fermion mass hierarchy is thus explained quite economically in our FN331-model, as the scalar content is sufficient to house the  FN-mechanism. The Froggatt-Nielsen mechanism also serves a purpose, other than that of generating the fermion mass hierarchy: the scalar mediated flavour changing neutral currents are in part suppressed due to the FN-mechanism. 

The traditional 331-models suffer from scalar mediated flavour changing neutral currents at tree-level, due to the  unequal treatment of quark generations, with no suppression mechanism. We find that the quark  flavour violating physical Higgs Yukawa couplings obtain suppression when the FN-charges of the quarks are such that the CKM-matrix texture is generated naturally.  The flavour violating quark Yukawa couplings are also suppressed by the scale of the $SU(3)_L\times U(1)_X$-breaking. We find that the scale of   the $SU(3)_L\times U(1)_X$-breaking has to be roughly  5 TeV or higher in order to satisfy the bounds coming from the quark flavour violation.  We provided three numerical examples showing that the scalar mediated flavour changing neutral currents are suppressed enough.

\vspace{0.5cm}

{\bf Acknowledgements.}  
KH and NK acknowledge the H2020-MSCA-RICE-2014 grant no. 645722 (NonMinimalHiggs). NK is supported by Vilho, Yrj{\"o} and Kalle V{\"a}is{\"a}l{\"a}
Foundation.

\appendix


\section{Gauge boson masses}\label{gauge boson masses appendix}

The covariant derivative is:
\be
D_{\mu}=\partial_{\mu}-ig_3 T_{a}W_{a\mu}-ig_{x}XB_{\mu}~~~~~~~~~~~~~~~~~~~~~~~~~~~~~~~~~~~~~~~~~~~~~~~~~~~~~~~~~~~~~~~~~~~~~~~~~~~~~~~~~~~~~~
\ee
\be\nonumber
\resizebox{1.0 \textwidth}{!} 
{
$ 
 =\partial_{\mu}-i\frac{g_3}{\sqrt{2}}\left(\begin{array}{ccc}
\frac{1}{\sqrt{2}}W_{3\mu}+\frac{1}{\sqrt{6}}W_{8\mu}+\frac{\sqrt{2}g_x}{g}XB_{\mu} & {W'}_{\mu}^{+} &  {X'}^{0}_{\mu}\\
 {W'}^{-}_{\mu} &-\frac{1}{\sqrt{2}}W_{3\mu}+\frac{1}{\sqrt{6}}W_{8\mu}+\frac{\sqrt{2}g_x}{g}XB_{\mu} &  {V'}^{-}_{\mu}\\
 X'^{0\ast}_{\mu} &  {V'}^{+}_{\mu}&-\frac{2}{\sqrt{6}}W_{8\mu}+\frac{\sqrt{2}g_x}{g}XB_{\mu}
\end{array}\right).
$
}
\ee 
The ${X'}^{0}_\mu$ is the non-hemitian neutral gauge boson:
\begin{equation}
{X'}^{0}_\mu=\frac{1}{\sqrt{2}}(W'_{4\mu}-iW'_{5\mu}).
\end{equation}
The mass matrices are as follows.

\subsection{Charged gauge bosons}
The charged gauge bosons mass term is given by,
\be
\mathcal{L}\supset Y_{charged}^T M^2_{charged} Y,
\ee
where $Y^T=({W'}_{\mu}^{+},{V'}_{\mu}^{+})$ and, 
\begin{equation}
M_{charged}^2=\left(\begin{array}{cc}
\frac{g_3^{2}}{4}({v'}^2+v_1^2)& \frac{g_3^{2}v_1 v_2}{4}\\
\frac{g_3^{2}v_1 v_2}{4} & \frac{g_3^{2}}{4}({v'}^2+v_2^2+u^2)
\end{array}\right).
\end{equation}
The eigenvalues of the matrix are:
\begin{eqnarray}
&&m_{W^{\pm}}^2
=  \frac{g_3^2}{4}({v'}^2+\frac{v_1^2 u^2}{v_2^2+u^2})+\mathcal{O}\left(\frac{v^2_{\textrm{light}}}{v^2_{\textrm{heavy}}}\right),\label{charged gauge boson masses}\\
&&m_{V^{\pm}}^2
= \frac{g_3^2}{4}(v_2^2+u^2)+\mathcal{O}\left(\frac{v^2_{\textrm{light}}}{v^2_{\textrm{heavy}}}\right).\nonumber
\end{eqnarray}
The mass eigenstates are defined as 
\begin{eqnarray}\label{tan of charged gauge boson mixing angle}
&&W^{-}_{\mu}=\cos\theta ~{W'}^{-}_{\mu}+\sin\theta ~{V'}^{-}_{\mu}\nonumber\\
&&V^{-}_{\mu}=-\sin\theta ~{W'}^{-}_{\mu}+\cos\theta ~{V'}^{-}_{\mu},\label{charged gauge boson mass eigenstates}
\end{eqnarray}
where the mixing angle $\theta$ is defined as:
\begin{equation}
\tan 2\theta=-\frac{2v_1 v_2}{v_2^2+u^2-v_1^2}.
\end{equation}
The Eq. (\ref{tan of charged gauge boson mixing angle}) leads to:
\begin{eqnarray}
&&\sin\theta = -\frac{v_1 v_2}{v_2^2+u^2}+\mathcal{O}\left(\frac{v^2_{\textrm{light}}}{v^2_{\textrm{heavy}}}\right),\nonumber\\
&&\cos\theta = 1+\mathcal{O}\left(\frac{v^2_{\textrm{light}}}{v^2_{\textrm{heavy}}}\right).\nonumber
\end{eqnarray}
The mixing angle is really small due to large difference in VEVs.
The  SM gauge boson $W^{\pm}_{\mu}$ is almost totally  ${W'}^{\pm}_{\mu}$ and  $V^{\pm}_{\mu}$ is mostly  ${V'}^{\pm}_{\mu}$.

\subsection{Neutral gauge bosons}
There are five neutral gauge bosons: $W_{3\mu}$, $W_{\mu}$, $B_\mu$, $W'_{4\mu}$ and $W_{5\mu}$. 
The imaginary part of   ${X'}^{0}_\mu$   decouples from the other neutral gauge bosons and acquires a mass,
\begin{equation}
M^2_{W_5}=\frac{g_3^2}{4}(v_1^2+v_2^2+u^2).
\end{equation}
We identify the state $W'_{5\mu}$ as mass eigenstate,
\begin{equation}
W'_{5\mu}\equiv W_{5\mu}.
\end{equation}

The neutral gauge bosons $W_{3\mu}$, $W_{\mu}$, $B_\mu$ and  $W'_{4\mu}$  mix,
\be
\mathcal{L}\supset \frac{1}{2} X^T M^2_{neutral} X,
\ee
where the basis is  $X^T=(W_{3\mu}, W_{8\mu},B_{\mu},W_{4\mu})$ and,
\begin{equation}
M^2_{neutral}=\frac{g_3^2}{4}\left(\begin{array}{cccc}
{v'}^2+v_1^2& \frac{v_1^2-{v'}^2}{\sqrt{3}} & -\frac{2 g_x}{3g_3}(v_1^2+2{v'}^2) &  v_1 v_2\\
\frac{v_1^2-{v'}^2}{\sqrt{3}}  &\frac{({v'}^2+v_1^2+4(v_2^2+u_2^2))}{3}&\frac{(-v_1^2+2(v_2^2+u_2^2+{v'}^2))}{ \left(\frac{3\sqrt{3}g_3}{2 g_x}\right)}  & -\frac{v_1 v_2}{\sqrt{3}}\\
 -\frac{2 g_x}{3g_3}(v_1^2+2{v'}^2) & \frac{(-v_1^2+2(v_2^2+u_2^2+{v'}^2))}{ (\frac{3\sqrt{3}g_3}{2 g_x})} & \frac{(v_1^2+v_2^2+u_2^2+4{v'}^2)}{(\frac{9g_3^2}{4g_x^{2}})} &  -\frac{4 g_x}{3g_3}(v_1 v_2)\\
 v_1 v_2& -\frac{v_1 v_2}{\sqrt{3}}  &  -\frac{4g_x }{3g_3}(v_1 v_2) & v_1^2+v_2^2+u_2^2
\end{array}\right).\nonumber
\end{equation}
The eigenvalues of this matrix can be solved analytically and they are:
\begin{eqnarray}
&&m_{\gamma}^2=0,\nonumber\\
&&m_{W_4}^2=\frac{g_3^2}{4}(u^2+v_2^2+v_1^2),\nonumber\\
&& m_Z^2 = \frac{g_3^2}{4}\left(\frac{3g_3^2+4g_x^2}{3g_3^2+g_x^2}\right)\left({v'}^2+\frac{v_1^2  u^2}{v_2^2+u^2}\right)+\mathcal{O}\left(\frac{v^2_{\textrm{light}}}{v^2_{\textrm{heavy}}}\right),
\nonumber\\
&& m_{Z'}^2 = \frac{3g_3^2+g_x^2}{9}(v_2^2+u^2)+\mathcal{O}\left(\frac{v^2_{\textrm{light}}}{v^2_{\textrm{heavy}}}\right).\nonumber
\end{eqnarray}
We notice  that one of the eigenvalues is exactly the same as that of the  imaginary part of the non-Hermitian gauge boson. We can therefore identify the combination:
\begin{equation}
X^0_\mu=\frac{1}{\sqrt{2}}(W_{4\mu}-iW_{5\mu})
\end{equation}
as the \emph{physical neutral non-hermitean gauge boson}.


\section{Charged currents}\label{charged currents}
Let us investigate the currents mediated by the charged gauge bosons $W_{\mu}^{\pm}$ and  $V_{\mu}^{\pm}$. The charged currents arrise from the kinetic terms of the fermions:
\begin{equation}\label{quark kinetic}
\mathcal{L}_{kin}=i\bar{Q}'_{L,1}\slashed D Q'_{L,1}+i\bar{Q}'_{L,2}\slashed D Q'_{L,2}+i\bar{Q}'_{L,3}\slashed D Q'_{L,3}.
\end{equation}
The covariant derivative acts differently on triplets and antitriplets:
\begin{eqnarray}
\textrm{triplet:}&&\quad D_{\mu}=\partial_{\mu}-i g_{3} T^{a}W^a_{\mu}-ig_{X} X B_{\mu}\nonumber\\
\textrm{antitriplet:}&&\quad D_{\mu}=\partial_{\mu}-i g_{3} \bar{T}^{a}W^a_{\mu}-ig_{X} X B_{\mu},\nonumber
\end{eqnarray}
where $\bar{T}^{a}=-(T^{a})^{\ast}$. Therefore:
\begin{equation}
-ig_3 T^{a}W_{\mu}^{a}=-i\frac{g_3}{2}\left(\begin{array}{ccc}
W_{3\mu}+\frac{1}{\sqrt{3}}W_{8\mu}& \sqrt{2}{W'}_{\mu}^{+} &  \sqrt{2}{X'}^{0}_{\mu}\\
 \sqrt{2}{W'}^{-}_{\mu} &-W_{3\mu}+\frac{1}{\sqrt{3}}W_{8\mu} &  \sqrt{2}{V'}^{-}_{\mu}\\
\sqrt{2} X'^{0\ast}_{\mu} &  \sqrt{2}{V'}^{+}_{\mu}&-\frac{2}{\sqrt{3}}W_{8\mu}
\end{array}\right),\nonumber
\end{equation}
and 
\begin{equation}
-ig_3 \bar{T}^{a}W_{\mu}^{a}=i\frac{g_3}{2}\left(\begin{array}{ccc}
W_{3\mu}+\frac{1}{\sqrt{3}}W_{8\mu}& \sqrt{2}{W'}_{\mu}^{-} &  \sqrt{2}{X'}^{0\ast}_{\mu}\\
 \sqrt{2}{W'}^{+}_{\mu} &-W_{3\mu}+\frac{1}{\sqrt{3}}W_{8\mu} &  \sqrt{2}{V'}^{+}_{\mu}\\
\sqrt{2} X'^{0}_{\mu} &  \sqrt{2}{V'}^{-}_{\mu}&-\frac{2}{\sqrt{3}}W_{8\mu}
\end{array}\right).\nonumber
\end{equation}

From the quark kinetic terms we acquire the $W_{\mu}^{\pm}$ and  $V_{\mu}^{\pm}$ interactions with quarks:

\begin{eqnarray}
\mathcal{L}_{gCC}&&=\frac{g_{3}}{\sqrt{2}}\Big[ (\bar{u}'_{1,L}~\bar{u}'_{2,L}~\bar{u}'_{3,L}~\bar{U}'_{1,L})\gamma^{\mu}
\left(\begin{array}{ccccc}
1 & 0 & 0 & 0 & 0\\
0 & 1 & 0 & 0 & 0\\
0 & 0 & 1 & 0 & 0\\
0 & 0 & 0 & 0 & 0
\end{array}\right)
\left(\begin{array}{c}
d'_{1,L} \\
d'_{2,L} \\
d'_{3,L} \\
D'_{1,L} \\
D'_{2,L}
\end{array}\right){W'}^{+}_{\mu}
+h.c.\Big]\nonumber\\
&&+\frac{g_{3}}{\sqrt{2}}\Big[ (\bar{u}'_{1,L}~\bar{u}'_{2,L}~\bar{u}'_{3,L}~\bar{U}'_{1,L})\gamma^{\mu}
\left(\begin{array}{ccccc}
0 & 0 & 0 & 0 & 0\\
0 & 0 & 0 & 1 & 0\\
0 & 0 & 0 & 0 & 1\\
1 & 0 & 0 & 0 & 0
\end{array}\right)
\left(\begin{array}{c}
d'_{1,L} \\
d'_{2,L} \\
d'_{3,L} \\
D'_{1,L} \\
D'_{2,L}
\end{array}\right){V'}^{+}_{\mu}
+h.c.\Big].\label{charged gauge boson couplings}
\end{eqnarray}
By writing this in terms of the gauge boson mass eigenstates, as defined in  Eq.  (\ref{charged gauge boson mass eigenstates}), we obtain:
\begin{eqnarray}
\mathcal{L}_{gCC}&&=\frac{g_{3}}{\sqrt{2}}\bar{u}'_L\gamma^{\mu}\Big[\cos\theta
\left(\begin{array}{ccccc}
1 & 0 & 0 & 0 & 0\\
0 & 1 & 0 & 0 & 0\\
0 & 0 & 1 & 0 & 0\\
0 & 0 & 0 & 0 & 0
\end{array}\right)
+\sin\theta
\left(\begin{array}{ccccc}
0 & 0 & 0 & 0 & 0\\
0 & 0 & 0 & 1 & 0\\
0 & 0 & 0 & 0 & 1\\
1 & 0 & 0 & 0 & 0
\end{array}\right)
\Big]d'_L {W}^{+}_{\mu}+h.c.\nonumber\\
&&+\frac{g_{3}}{\sqrt{2}}\bar{u}'_L\gamma^{\mu}\Big[\cos\theta
\left(\begin{array}{ccccc}
0 & 0 & 0 & 0 & 0\\
0 & 0 & 0 & 1 & 0\\
0 & 0 & 0 & 0 & 1\\
1 & 0 & 0 & 0 & 0
\end{array}\right)
-\sin\theta
\left(\begin{array}{ccccc}
1 & 0 & 0 & 0 & 0\\
0 & 1 & 0 & 0 & 0\\
0 & 0 & 1 & 0 & 0\\
0 & 0 & 0 & 0 & 0
\end{array}\right)
\Big]d'_L {V}^{+}_{\mu}+h.c.\nonumber
\end{eqnarray}

\section{Additional contributions to the quark Yukawa-interactions}\label{quark FN extra contribution}
The complete set of  Yukawa couplings of Higgs to quarks were given as:
\begin{eqnarray}
\mathcal{L}_{quark Yukawa}&=& \frac{1}{\sqrt{2}}\bar{u}_{L}U_L^u({\Gamma'}^u_{h}+{\Delta'}^u_{h})U_{R}^{u\dagger}u_{R}~h+\frac{1}{\sqrt{2}}\bar{d}_{L}U_{L}^d({\Gamma'}^d_{h}+{\Delta'}^d_{h})U_{R}^{d\dagger}d'_{R}~h,\nonumber
\end{eqnarray}
where the ${\Gamma'}^u_h$ and ${\Gamma'}^d_h$  are the dominant contribution to the SM-quark Yukawa couplings discussed in detail in Section \ref{Higgs mediated neutral meson mixing}. Here we will give also the sub-dominant FN-contribution for completeness:
\begin{eqnarray}
{\Delta'}^u_h&=&\widetilde{\Delta}^u\left[\frac{U^H_{42}}{v_2}+\frac{U^H_{52}}{u}+\frac{v_1}{v_2 u}U^H_{32}\right],\nonumber\\
{\Delta'}^d_h&=&\widetilde{\Delta}^d\left[\frac{U^H_{42}}{v_2}+\frac{U^H_{52}}{u}+\frac{v_1}{v_2 u}U^H_{32}\right],\nonumber
\end{eqnarray}
where the $\widetilde{\Delta}^u$ and $\widetilde{\Delta}^d$ are given in the as,
\begin{eqnarray}
\resizebox{0.92 \textwidth}{!} 
{
$ 
\widetilde{\Delta}^{u}=\left(\begin{array}{cccc}
k^\rho_{11}m^{u}_{11} &
 k^\rho_{12}m^{u}_{12} &
k^\rho_{13}m^{u}_{13} &
 k^\rho_{14}m^{u}_{14} \\ 
k^{\eta^\ast}_{21}m^{u}_{21}&
 k^{\eta^\ast}_{22}m^{u}_{22} &
 k^{\eta^\ast}_{23}m^{u}_{23}&
 k^{\eta^\ast}_{24}m^{u}_{24} \\ 
k^{\eta^\ast}_{31}m^{u}_{31} & 
k^{\eta^\ast}_{32}m^{u}_{32} &
 k^{\eta^\ast}_{33}m^{u}_{33}&
 k^{\eta^\ast}_{34}m^{u}_{34} \\ 
k^\rho_{11}m^{u}_{41}&
 k^\rho_{12}m^{u}_{42}&
 k^\rho_{13}m^{u}_{43} &
 k^\rho_{14}m^{u}_{44} \\ 
\end{array}\right)+u\left[q(\chi)-q(\rho)\right]\left(\begin{array}{cccc}
0 &
0 &
0 & 
0  \\ 
0 &
0 & 
 0 &
0  \\ 
0 & 
0 &
0 &
 0 \\ 
(y_{\chi}^{u})_{11}&
(y_{\chi}^{u})_{12}&
  (y_{\chi}^{u})_{13} &
(y_{\chi}^{u})_{14} \\ 
\end{array}\right),\label{extra contribution to up Yukawas from FN}
$
}
\end{eqnarray}
and,
\begin{eqnarray}
&&\widetilde{\Delta}^{d}=\left(\begin{array}{ccccc}
h^\eta_{11}(m^{d})_{11} & 
h^\eta_{12}(m^{d})_{12} & 
 h^\eta_{13}(m^{d})_{13} &
 h^\eta_{14}(m^{d})_{14}&
h^\eta_{15}(m^{d})_{15}\\ 
h^{\rho^\ast}_{21}(m^{d})_{21} &
 h^{\rho^\ast}_{22}(m^{d})_{22}&
h^{\rho^\ast}_{23}(m^{d})_{23}&
h^{\rho^\ast}_{24}(m^{d})_{24}&
h^{\rho^\ast}_{25}(m^{d})_{25}\\ 
h^{\rho^\ast}_{31}(m^{d})_{31}&
h^{\rho^\ast}_{32}(m^{d})_{32}&
h^{\rho^\ast}_{33}(m^{d})_{33}&
h^{\rho^\ast}_{34}(m^{d})_{34}&
h^{\rho^\ast}_{31}(m^{d})_{31}\\ 
h^{\rho^\ast}_{21}(m^{d})_{21}&
 h^{\rho^\ast}_{22}(m^{d})_{22}&
 h^{\rho^\ast}_{23}(m^{d})_{23} &
 h^{\rho^\ast}_{24}(m^{d})_{24}&
h^{\rho^\ast}_{25}(m^{d})_{25}\\ 
h^{\rho^\ast}_{31}(m^{d})_{31}&
 h^{\rho^\ast}_{32}(m^{d})_{32}&
 h^{\rho^\ast}_{33}(m^{d})_{33} &
 h^{\rho^\ast}_{34}(m^{d})_{34}&
h^{\rho^\ast}_{35}(m^{d})_{35}\\ 
\end{array}\right)\nonumber\\
&&+u\left[q(\chi^\ast)-q(\rho^\ast)\right]\left(\begin{array}{ccccc}
0 & 
0& 
0 &
 0&
0\\ 
0 &
 0&
0&
0&
0\\ 
0&
0&
0&
0&
0\\ 
(y_{{\chi^\ast}}^{d})_{21}&
 (y_{{\chi^\ast}}^{d})_{22}&
(y_{{\chi^\ast}}^{d})_{23} &
(y_{{\chi^\ast}}^{d})_{24}&
(y_{{\chi^\ast}}^{d})_{25}\\ 
(y_{{\chi^\ast}}^{d})_{31}&
 (y_{{\chi^\ast}}^{d})_{32}&
 (y_{{\chi^\ast}}^{d})_{33} &
 (y_{{\chi^\ast}}^{d})_{34}&
(y_{{\chi^\ast}}^{d})_{35}\\ 
\end{array}\right).\label{extra contribution to down Yukawas from FN}
\end{eqnarray}
The additional physical contribution to Higgs Yukawa couplings from Froggatt-Nielsen mechanism is:
\begin{eqnarray}
&&(\Delta^{u}_h)_{ij}=\left\{ a^{u}_k(U^{u}_L)_{ik}(U^{u\dagger}_L)_{kj}m^{u}_j+ b^{u}_k(U^{u}_R)_{ik}(U^{u\dagger}_R)_{kj}m_i^{u}
\right\}\left(\frac{U^H_{42}}{v_2}+\frac{U^H_{52}}{u}+\frac{v_1}{v_2 u}U^{H}_{32}\right)\nonumber\\
&&+\sqrt{2}m_j\left[q(\chi)-q(\rho)\right]\left[(U^{u}_L)_{i4}(U^{u\dagger}_R)_{4j}
-\frac{v_2}{v_1}(U^{u}_L)_{i4}(U^{u\dagger}_R)_{1j}\right]\left(\frac{U^H_{42}}{v_2}+\frac{U^H_{52}}{u}+\frac{v_1}{v_2 u}U^{H}_{32}\right),\nonumber
\end{eqnarray}
where,
\begin{eqnarray}
&&a^{u}_1=a^{u}_4=q(\bar{Q}_{L,1})+q(\rho), \quad a^{u}_{2}=q(\bar{Q}_{L,2})+q(\eta^\ast),\nonumber\\
&&a^{u}_{3}=q(\bar{Q}_{L,3})+q(\eta^\ast), \quad b^{u}_j=q(u_{R,j}).\nonumber
\end{eqnarray}
\begin{eqnarray}
(\Delta^{d}_h)_{ij}&=&\left\{m^d_j (U^d_L)_{ik}(U_L^{d\dagger})_{kj}a_k^{d}
+m^d_i (U^d_R)_{ik}(U_R^{d\dagger})_{kj}b_k^{d}\right\}\left(\frac{U^H_{42}}{v_2}+\frac{U^H_{52}}{u}+\frac{v_1}{v_2 u}U^{H}_{32}\right)\nonumber\\
&&+\sqrt{2}m_j\left[q(\chi^\ast)-q(\rho^\ast)\right]\left[(U^d_L)_{i4}(U^{d\dagger})_R)_{4j}
+(U^d_L)_{i5}(U^{d\dagger})_R)_{5j}\right.\nonumber\\
&&\left.-\frac{v_2}{v_1}(U^d_L)_{i4}(U^{d\dagger})_R)_{2j}
-\frac{v_2}{v_1}(U^d_L)_{i5}(U^{d\dagger})_R)_{3j}\right]\left(\frac{U^H_{42}}{v_2}+\frac{U^H_{52}}{u}+\frac{v_1}{v_2 u}U^{H}_{32}\right),\nonumber
\end{eqnarray}
where,
\begin{eqnarray}
&&a^{d}_1=q(\bar{Q}_{L,1})+q(\eta), \quad a^{d}_{2}=a^d_4=q(\bar{Q}_{L,2})+q(\rho^\ast),\nonumber\\
&&a^{d}_{3}=a^d_5=q(\bar{Q}_{L,3})+q(\rho^\ast), \quad b_j=q(d_{R,j}).\nonumber
\end{eqnarray}
Notice that in both $(\Delta^{u}_h)_{ij}$ and $(\Delta^{d}_h)_{ij}$  the  couplings of SM quarks amongst themselves are heavily suppressed since the SM quarks are much lighter than the $SU(3)_L$-breaking scale.

\section{Higgs mediated neutral meson mixing at 1-loop level}\label{neutral meson mixing at 1 loop}
The heavy quarks enter the neutral meson mixing through diagrams of the form of Figure \ref{meson mixing loop}. To avoid clutter, we only take into account these diagrams, as they have the same fermion in both virtual lines.  

\begin{minipage}{\linewidth}
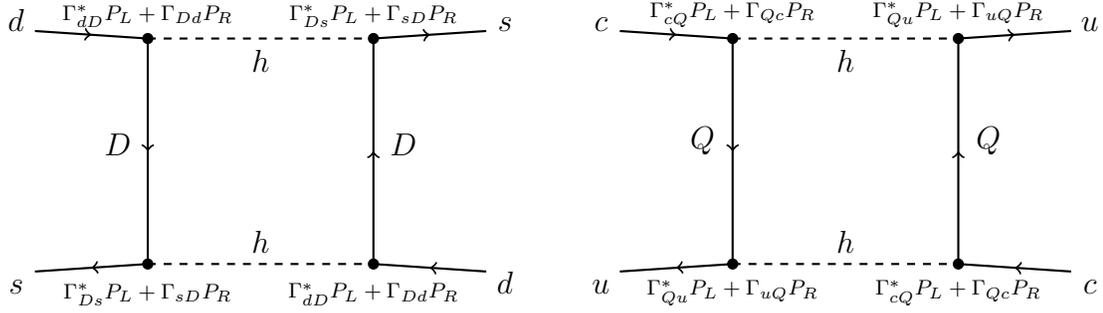
\begin{figure}[H]
\begin{center}
\begin{tikzpicture}[thick,scale=1.0]
\draw[particle] (-1.5,0.1) -- node[black,above,xshift=-1.0cm,yshift=-0.1cm] {$d$} (0,0);
\fill[black] (0,0) circle (0.06cm);
\draw (0,0) circle (0.06cm);
\draw (-0.01,0) -- node[black,above,xshift=-0.0cm,yshift=0.0cm] {\scriptsize $\Gamma^\ast_{dD}P_L+\Gamma_{Dd}P_R$} (0.01,0);
\draw[dashed] (0,0) -- node[black,above,yshift=-0.6cm,xshift=-0.0cm] {$h$} (3,0);
\fill[black] (3,0) circle (0.06cm);
\draw (3,0) circle (0.06cm);
\draw (2.99,0) -- node[black,above,xshift=-0.0cm,yshift=0.0cm] {\scriptsize $\Gamma^\ast_{Ds}P_L+\Gamma_{sD}P_R$} (3.01,0);
\draw[particle] (3,0) -- node[black,above,yshift=-0.1cm,xshift=1.0cm] {$s$} (4.5,0.1);
\draw[particle] (0,0) -- node[black,above,xshift=-0.4cm,yshift=-0.2cm] {$D$} (0,-3);
\fill[black] (0,-3) circle (0.06cm);
\draw (0,-3) circle (0.06cm);
\draw (-0.01,-3) -- node[black,above,xshift=-0.0cm,yshift=-0.7cm] {\scriptsize $\Gamma^\ast_{Ds}P_L+\Gamma_{sD}P_R$} (0.01,-3);
\draw[dashed] (0,-3) -- node[black,above,yshift=0.0cm,xshift=-0.0cm] {$h$} (3,-3);
\draw[particle] (0,-3) -- node[black,above,xshift=-1.0cm,yshift=-0.5cm] {$s$} (-1.5,-3.1);
\draw[particle] (4.5,-3.1) -- node[black,above,yshift=-0.5cm,xshift=1.0cm] {$d$} (3,-3);
\fill[black] (3,-3) circle (0.06cm);
\draw (3,-3) circle (0.06cm);
\draw (2.99,-3) -- node[black,above,xshift=-0.0cm,yshift=-0.7cm] {\scriptsize $\Gamma^\ast_{dD}P_L+\Gamma_{Dd}P_R$} (3.01,-3);
\draw[particle] (3,-3) -- node[black,above,yshift=-0.2cm,xshift=0.4cm] {$D$} (3,0);
\end{tikzpicture}
\hspace{5mm}
\begin{tikzpicture}[thick,scale=1.0]
\draw[particle] (-1.5,0.1) -- node[black,above,xshift=-1.0cm,yshift=-0.1cm] {$c$} (0,0);
\fill[black] (0,0) circle (0.06cm);
\draw (0,0) circle (0.06cm);
\draw (-0.01,0) -- node[black,above,xshift=-0.0cm,yshift=0.0cm] {\scriptsize $\Gamma^\ast_{cQ}P_L+\Gamma_{Qc}P_R$} (0.01,0);
\draw[dashed] (0,0) -- node[black,above,yshift=-0.6cm,xshift=-0.0cm] {$h$} (3,0);
\fill[black] (3,0) circle (0.06cm);
\draw (3,0) circle (0.06cm);
\draw (2.99,0) -- node[black,above,xshift=-0.0cm,yshift=0.0cm] {\scriptsize $\Gamma^\ast_{Qu}P_L+\Gamma_{uQ}P_R$} (3.01,0);
\draw[particle] (3,0) -- node[black,above,yshift=-0.1cm,xshift=1.0cm] {$u$} (4.5,0.1);
\draw[particle] (0,0) -- node[black,above,xshift=-0.4cm,yshift=-0.2cm] {$Q$} (0,-3);
\fill[black] (0,-3) circle (0.06cm);
\draw (0,-3) circle (0.06cm);
\draw (-0.01,-3) -- node[black,above,xshift=-0.0cm,yshift=-0.7cm] {\scriptsize $\Gamma^\ast_{Qu}P_L+\Gamma_{uQ}P_R$} (0.01,-3);
\draw[dashed] (0,-3) -- node[black,above,yshift=0.0cm,xshift=-0.0cm] {$h$} (3,-3);
\draw[particle] (0,-3) -- node[black,above,xshift=-1.0cm,yshift=-0.5cm] {$u$} (-1.5,-3.1);
\draw[particle] (4.5,-3.1) -- node[black,above,yshift=-0.5cm,xshift=1.0cm] {$c$} (3,-3);
\fill[black] (3,-3) circle (0.06cm);
\draw (3,-3) circle (0.06cm);
\draw (2.99,-3) -- node[black,above,xshift=-0.0cm,yshift=-0.7cm] {\scriptsize $\Gamma^\ast_{cQ}P_L+\Gamma_{Qc}P_R$} (3.01,-3);
\draw[particle] (3,-3) -- node[black,above,yshift=-0.2cm,xshift=0.4cm] {$Q                                                                                                                                                                                                                                                                                                                                                                                                                                                                                                                                                                                                                                                                                                                                                                                                                                                                                                                                       $} (3,0);
\end{tikzpicture}
\vspace{0.5cm}
\caption{Neutral meson mixing diagrams mediated by heavy quarks: $K^0$-$\bar{K}^0$ (left) and $D^0$-$\bar{D}^0$ (right). The $D=D_1,D_2$ on the left and $Q=t,U$ on the right.~~~~~~~~~~~~~~~~~~~~~~~~~~~~}
\label{meson mixing loop} 
\end{center}
\end{figure}    
\end{minipage}

We derive bounds on Yukawas involving heavy quarks by assuming that the diagrams in Figure \ref{meson mixing loop} receive contribution only from one type of virtual quark. 
The Wilson coefficients from  $D^0$-$\bar{D}^0$ are: 

\begin{eqnarray}
C^{cu}_1 & = & \frac{1}{16}\frac{1}{16 \pi^2}\frac{S_1^{H}(x_{Qh})}{m_h^2}
\left[(\Gamma^u_h)_{uQ}(\Gamma^{u\ast}_h)_{cQ}\right]^2,\label{exotic up wilson}\\
\widetilde{C}^{cu}_1 &  = & \frac{1}{16}\frac{1}{16 \pi^2}\frac{S_1^{H}(x_{Qh})}{m_h^2}
\left[(\Gamma^{u\ast}_h)_{Qu}(\Gamma^{u}_h)_{Qc}\right]^2,\nonumber\\
C^{cu}_2 & = & -\frac{1}{16}\frac{1}{16 \pi^2}\frac{S_2^{H}(x_{Qh})}{m_h^2}
\left[(\Gamma^{u\ast}_h)_{Qu}(\Gamma^{u\ast}_h)_{cQ}\right]^2,\nonumber\\
\widetilde{C}^{cu}_2 & = & -\frac{1}{16}\frac{1}{16 \pi^2}\frac{S_2^{H}(x_{Qh})}{m_h^2}
\left[(\Gamma^{u}_h)_{uQ}(\Gamma^{u}_h)_{Qc}\right]^2,\nonumber\\
C^{cu}_4 & = & -\frac{1}{8}\frac{1}{16 \pi^2}\frac{S_2^{H}(x_{Qh})}{m_h^2}
(\Gamma^{u\ast}_h)_{Qu}(\Gamma^{u\ast}_h)_{cQ}(\Gamma^{u}_h)_{uQ}(\Gamma^{u}_h)_{Qc},\nonumber\\
\widetilde{C}^{cu}_4 & = & -\frac{1}{8}\frac{1}{16 \pi^2}\frac{S_2^{H}(x_{Qh})}{m_h^2}
(\Gamma^{u}_h)_{uQ}(\Gamma^{u}_h)_{Qc}(\Gamma^{u\ast}_h)_{Qu}(\Gamma^{u\ast}_h)_{cQ},\nonumber\\
C^{cu}_5 & = & -\frac{1}{4}\frac{1}{16 \pi^2}\frac{S_1^{H}(x_{Qh})}{m_h^2}
(\Gamma^{u\ast}_h)_{Qu}(\Gamma^{u}_h)_{Qc}(\Gamma^{u}_h)_{uQ}(\Gamma^{u\ast}_h)_{cQ},\nonumber\\
\widetilde{C}^{cu}_5 & = & -\frac{1}{4}\frac{1}{16 \pi^2}\frac{S_1^{H}(x_{Qh})}{m_h^2}
(\Gamma^{u}_h)_{uQ}(\Gamma^{u\ast}_h)_{cQ}(\Gamma^{u\ast}_h)_{Qu}(\Gamma^{u}_h)_{Qc},\nonumber
\end{eqnarray}
where,
\begin{eqnarray}
S_1^{H}(x_{Qh}) &=& \frac{x^2-1-2x\log x}{2 (x-1)^3},\quad S_2^{H}(x_{Qh}) = \frac{2x(2-2x+(1+x)\log x)}{ (x-1)^3},\nonumber
\end{eqnarray}
with $x_{Qt}=m_Q^2/m_h^2$. The Wilson coefficients for $K^0$-$\bar{K}^0$, $B_d^0$-$\bar{B}_d^0$ and $B_s^0$-$\bar{B}_s^0$  are obtained from Eq.  (\ref{exotic up wilson}) by appropriate flavour index replacement. 

We obtain bounds on Tables \ref{quark flavour bounds2} and \ref{exotic quark flavour bounds}  by demanding that the each operator  must individually satisfy the flavour constraint. The Wilson coefficient $C_1$ for a given meson is the dominant one-loop contribution. This is the only Wilson coefficient whose  Yukawa-couplings are all  proportional to exotic quark mass.

\section{Top quark decay}\label{top quark decay appendix}
\subsection{$t\to hc$}
The decay  $t\to h c$ occurs at tree-level in our model. One can therefore place a bound directly on the flavour violating Yukawa coupling. The bound from this process however is quite weak as seen in the Table \ref{quark flavour bounds2}.
The rate for flavour violating top decay $t\to h c$ at tree-level is \cite{Casagrande:2008hr}:

\begin{equation}
BR(t\to h c)=\frac{(m_t^2-m_h^2)^2 m_W^2 m_t^2}{g^2(m_t^2-m_W^2)^2(m_t^2+2m_W^2)^2}\left(\lvert \Gamma^u_{ct}\rvert^2+\lvert \Gamma^u_{tc}\rvert^2+\frac{4m_t  m_c}{m_t^2-m_h^2}Re\lvert \Gamma^u_{ct}\Gamma^u_{tc}\rvert\right).
\end{equation}

The current experimental bounds are: $BR(t\to hc)<0.56\%$ (CMS) \cite{Khachatryan:2014jya}
 and $BR(t\to hu)<0.79\%$ (ATLAS) \cite{Aad:2014dya}.

\subsection{$t\to q\gamma$}
The branching ratio for $t\to q\gamma$ is given by,
\begin{equation}
\textrm{BR}(t\to q\gamma)=\frac{m_t^3}{4\pi}\left(\lvert A^q_L\rvert+\lvert A^q_R\rvert\right),
\end{equation}
where,
\begin{equation}
A_R^q=-\frac{Q_u e}{32\pi^2}\sum_{i=u,c,t,U}\frac{(\Gamma^u_h)_{qi}}{2}
\int^1_0 dx\int^x_0 dy\frac{(\Gamma^{u\ast}_h)_{ti}y(y-x)m_t+(\Gamma^{u}_h)_{it}(y-1)m_i}{y(y-x)m_t^2+(1-y)m_i^2+ym_h^2},
\end{equation}
\begin{equation}
A_L^q=-\frac{Q_u e}{32\pi^2}\sum_{i=u,c,t,U}\frac{(\Gamma^{u\ast}_h)_{iq}}{2}
\int^1_0 dx\int^x_0 dy\frac{(\Gamma^{u}_h)_{it}y(y-x)m_t+(\Gamma^{u\ast}_h)_{ti}(y-1)m_i}{y(y-x)m_t^2+(1-y)m_i^2+ym_h^2},
\end{equation}
and $q=u,c$.

The experimental bound for branching ratio of $t\to q\gamma$ is \cite{Patrignani:2016xqp}:
\begin{equation}
\textrm{Br}(t\to q\gamma)<5.9\times 10^{-3},
\end{equation}
with $q=u,c$. The bound from this process is so weak that it is automatically satisfied by our model.

\section{Numerical example}\label{numerical example}
\subsection{Example 1}\label{numerical example 1}

 The $SU(3)_L\times U(1)_X$-breaking VEVs are set to be $u=48000$ GeV and $v_2=55000$ GeV, and the $SU(2)_L\times U(1)_Y$-breaking VEVs are $v_1=100$ GeV and $v'=237.05$ GeV.
 The scalar potential parameters used are: $\lambda_1=0.4$, $\lambda_2=0.2898$, $\lambda_3=0.5$, $\widetilde{\lambda}_{12}=0.2$, $\widetilde{\lambda}_{13}=0.5$, $\widetilde{\lambda}_{23}=-0.6$, $\lambda_{12}=0.2$, $\lambda_{13}=0.45159$, $\lambda_{23}=0.22$,  and $b=-(10000~~\textrm{GeV})^2$.
All the scalar masses are positive with these parameters.
The order-one numbers used are:
\begin{eqnarray}
\left(
\begin{array}{ccccc}
(c_{\eta}^d)_{11} & 
(c_{\eta}^d)_{12} &
(c_{\eta}^d)_{13} & 
(c_{\eta}^d)_{14}  \\
(c_{\rho^\ast}^d)_{21} &
(c_{\rho^\ast}^d)_{22} &
(c_{\rho^\ast}^d)_{23} & 
(c_{\rho^\ast}^d)_{24} \\
(c_{\rho^\ast}^d)_{31} & 
(c_{\rho^\ast}^d)_{32} &
(c_{\rho^\ast}^d)_{33} &
(c_{\rho^\ast}^d)_{34} \\
(c_{\chi^\ast}^d)_{21} &
(c_{\chi^\ast}^d)_{22} &
(c_{\chi^\ast}^d)_{23} &
(c_{\chi^\ast}^d)_{24}  \\
\end{array}
\right)
=
\left(
\begin{array}{ccccc}
-1.4451 &      	
2.4758 &			
1.1457  & 			
1.5480  &	 \\		
-0.6825&			
-1.5997  &		
1.0563  &		
1.4025  \\	
1.3606 & 	
1.3458  &		
0.9191 &		
1.7592 \\	
1.6103  &		
0.6771 &		
2.0789  &		
1.0260  \\	
\end{array}
\right),
\nonumber
\end{eqnarray}
and,
\begin{eqnarray}
\resizebox{1.0 \textwidth}{!} 
{
$ 
 \left(
\begin{array}{ccccc}
(c_{\eta}^d)_{11} & 
(c_{\eta}^d)_{12} &
(c_{\eta}^d)_{13} & 
(c_{\eta}^d)_{14} &
(c_{\eta}^d)_{15}  \\
(c_{\rho^\ast}^d)_{21} &
(c_{\rho^\ast}^d)_{22} &
(c_{\rho^\ast}^d)_{23} & 
(c_{\rho^\ast}^d)_{24} &
(c_{\rho^\ast}^d)_{25}  \\
(c_{\rho^\ast}^d)_{31} & 
(c_{\rho^\ast}^d)_{32} &
(c_{\rho^\ast}^d)_{33} &
(c_{\rho^\ast}^d)_{34} &
(c_{\rho^\ast}^d)_{35}  \\
(c_{\chi^\ast}^d)_{21} &
(c_{\chi^\ast}^d)_{22} &
(c_{\chi^\ast}^d)_{23} &
(c_{\chi^\ast}^d)_{24} &
(c_{\chi^\ast}^d)_{25}  \\
(c_{\chi^\ast}^d)_{31} &
(c_{\chi^\ast}^d)_{32} & 
(c_{\chi^\ast}^d)_{33} & 
(c_{\chi^\ast}^d)_{34} &
(c_{\chi^\ast}^d)_{35}  \\
\end{array}
\right)
=
\left(
\begin{array}{ccccc}
3.1761 &      	
1.0211 &			
1.3213 & 			
1.0249 &			
-1.0774  \\		
-1.3511 &			
2.0779 &		
-0.8241 &		
1.3531 &		
1.7105  \\	
-1.1187 & 	
1.9181 &		
1.9365 &		
1.8366 &		
-0.5927  \\	
0.9726 &		
-0.9809 &		
-1.5498 &		
4.4029 &		
-4.7242  \\	
1.7293 &		
-1.0878 & 	
-0.5440 & 	
-4.5289 &		
-3.4392  \\	
\end{array}
\right).
\nonumber
$
}
\end{eqnarray}

\subsection{Example 2}\label{numerical example 2}

 The $SU(3)_L\times U(1)_X$-breaking VEVs are set to be $u=21000$ GeV and $v_2=19000$ GeV, and the $SU(2)_L\times U(1)_Y$-breaking VEVs are $v_1=197.5999$ GeV and $v'=197.5999$ GeV.
 The scalar potential parameters used are: $\lambda_1=0.4$, $\lambda_2=0.2898$, $\lambda_3=0.8$, $\widetilde{\lambda}_{12}=0.4$, $\widetilde{\lambda}_{13}=0.5$, $\widetilde{\lambda}_{23}=-0.1$, $\lambda_{12}=0.5$, $\lambda_{13}=0.34165$, $\lambda_{23}=0.22$,  and $b=-(5600~~\textrm{GeV})^2$.
All the scalar masses are positive with these parameters.
The order-one numbers used are:
\begin{eqnarray}
\left(
\begin{array}{ccccc}
(c_{\eta}^d)_{11} & 
(c_{\eta}^d)_{12} &
(c_{\eta}^d)_{13} & 
(c_{\eta}^d)_{14}  \\
(c_{\rho^\ast}^d)_{21} &
(c_{\rho^\ast}^d)_{22} &
(c_{\rho^\ast}^d)_{23} & 
(c_{\rho^\ast}^d)_{24} \\
(c_{\rho^\ast}^d)_{31} & 
(c_{\rho^\ast}^d)_{32} &
(c_{\rho^\ast}^d)_{33} &
(c_{\rho^\ast}^d)_{34} \\
(c_{\chi^\ast}^d)_{21} &
(c_{\chi^\ast}^d)_{22} &
(c_{\chi^\ast}^d)_{23} &
(c_{\chi^\ast}^d)_{24}  \\
\end{array}
\right)
=
\left(
\begin{array}{ccccc}
3.0322 &      	
2.0844 &			
1.1522 & 			
1.6286 &	 \\		
-0.8684 &			
1.1254   &		
1.6668   &		
1.3133  \\	
1.3702 & 	
1.8171  &		
-0.7254 &		
1.9335 \\	
1.6895  &		
0.6371 &		
1.8873  &		
-1.7056  \\	
\end{array}
\right),
\nonumber
\end{eqnarray}
and,
\begin{eqnarray}
\resizebox{1.0 \textwidth}{!} 
{
$ 
 \left(
\begin{array}{ccccc}
(c_{\eta}^d)_{11} & 
(c_{\eta}^d)_{12} &
(c_{\eta}^d)_{13} & 
(c_{\eta}^d)_{14} &
(c_{\eta}^d)_{15}  \\
(c_{\rho^\ast}^d)_{21} &
(c_{\rho^\ast}^d)_{22} &
(c_{\rho^\ast}^d)_{23} & 
(c_{\rho^\ast}^d)_{24} &
(c_{\rho^\ast}^d)_{25}  \\
(c_{\rho^\ast}^d)_{31} & 
(c_{\rho^\ast}^d)_{32} &
(c_{\rho^\ast}^d)_{33} &
(c_{\rho^\ast}^d)_{34} &
(c_{\rho^\ast}^d)_{35}  \\
(c_{\chi^\ast}^d)_{21} &
(c_{\chi^\ast}^d)_{22} &
(c_{\chi^\ast}^d)_{23} &
(c_{\chi^\ast}^d)_{24} &
(c_{\chi^\ast}^d)_{25}  \\
(c_{\chi^\ast}^d)_{31} &
(c_{\chi^\ast}^d)_{32} & 
(c_{\chi^\ast}^d)_{33} & 
(c_{\chi^\ast}^d)_{34} &
(c_{\chi^\ast}^d)_{35}  \\
\end{array}
\right)
=
\left(
\begin{array}{ccccc}
3.7669 &      	
1.0530 &			
0.7468 & 			
0.7758 &			
-1.7566  \\		
-1.1387 &			
1.2912 &		
-0.5847 &		
1.9404 &		
2.0690  \\	
-1.3395 & 	
0.7667 &		
1.6400 &		
2.2566 &		
-0.9852  \\	
1.2627 &		
-0.9742 &		
-0.8593 &		
4.6157 &		
-4.4440 \\	
1.8535 &		
-0.8220 & 	
-0.7183 & 	
-4.0511 &		
-1.6860  \\
\end{array}
\right).
\nonumber
$
}
\end{eqnarray}

\subsection{Example 3}\label{numerical example 3}

 The $SU(3)_L\times U(1)_X$-breaking VEVs are set to be $u=5000$ GeV and $v_2=5500$ GeV, and the $SU(2)_L\times U(1)_Y$-breaking VEVs are $v_1=204.117$ GeV and $v'=204.117$ GeV.
 The scalar potential parameters used are: $\lambda_1=0.4$, $\lambda_2=0.4$, $\lambda_3=0.2$, $\widetilde{\lambda}_{12}=0.2$, $\widetilde{\lambda}_{13}=0.5$, $\widetilde{\lambda}_{23}=-0.6$, $\lambda_{12}=0.2$, $\lambda_{13}=0.13553$, $\lambda_{23}=0.22$,  and $b=-(3600~~\textrm{GeV})^2$.
All the scalar masses are positive with these parameters.
The order-one numbers used are:
\begin{eqnarray}
\left(
\begin{array}{ccccc}
(c_{\eta}^d)_{11} & 
(c_{\eta}^d)_{12} &
(c_{\eta}^d)_{13} & 
(c_{\eta}^d)_{14}  \\
(c_{\rho^\ast}^d)_{21} &
(c_{\rho^\ast}^d)_{22} &
(c_{\rho^\ast}^d)_{23} & 
(c_{\rho^\ast}^d)_{24} \\
(c_{\rho^\ast}^d)_{31} & 
(c_{\rho^\ast}^d)_{32} &
(c_{\rho^\ast}^d)_{33} &
(c_{\rho^\ast}^d)_{34} \\
(c_{\chi^\ast}^d)_{21} &
(c_{\chi^\ast}^d)_{22} &
(c_{\chi^\ast}^d)_{23} &
(c_{\chi^\ast}^d)_{24}  \\
\end{array}
\right)
=
\left(
\begin{array}{ccccc}
2.7513 &      	
1.9544 &			
1.2750  & 			
1.7015  &	 \\		
0.7884 &			
3.9203  &		
2.1641  &		
1.4914  \\	
0.8600 & 	
2.0938  &		
1.2073 &		
1.8613\\	
1.6398  &		
0.9957 &		
2.1263  &		
0.9763 \\	
\end{array}
\right),
\nonumber
\end{eqnarray}
and,
\begin{eqnarray}
\resizebox{1.0 \textwidth}{!} 
{
$ 
 \left(
\begin{array}{ccccc}
(c_{\eta}^d)_{11} & 
(c_{\eta}^d)_{12} &
(c_{\eta}^d)_{13} & 
(c_{\eta}^d)_{14} &
(c_{\eta}^d)_{15}  \\
(c_{\rho^\ast}^d)_{21} &
(c_{\rho^\ast}^d)_{22} &
(c_{\rho^\ast}^d)_{23} & 
(c_{\rho^\ast}^d)_{24} &
(c_{\rho^\ast}^d)_{25}  \\
(c_{\rho^\ast}^d)_{31} & 
(c_{\rho^\ast}^d)_{32} &
(c_{\rho^\ast}^d)_{33} &
(c_{\rho^\ast}^d)_{34} &
(c_{\rho^\ast}^d)_{35}  \\
(c_{\chi^\ast}^d)_{21} &
(c_{\chi^\ast}^d)_{22} &
(c_{\chi^\ast}^d)_{23} &
(c_{\chi^\ast}^d)_{24} &
(c_{\chi^\ast}^d)_{25}  \\
(c_{\chi^\ast}^d)_{31} &
(c_{\chi^\ast}^d)_{32} & 
(c_{\chi^\ast}^d)_{33} & 
(c_{\chi^\ast}^d)_{34} &
(c_{\chi^\ast}^d)_{35}  \\
\end{array}
\right)
=
\left(
\begin{array}{ccccc}
-0.5726 &      	
2.8213 &			
1.7018 & 			
0.6296 &			
-0.8344  \\		
-1.6252 &			
2.8103 &		
1.9561 &		
2.1157 &		
-2.5466  \\	
2.3977 & 	
1.1451 &		
0.5187 &		
-0.9436 &		
1.5133  \\	
-1.7660 &		
-1.3697 &		
0.9654 &		
-0.9014 &		
-1.3291  \\	
2.6824 &		
2.8886 & 	
-2.0420 & 	
1.4547 &	
1.7573  \\	
\end{array}
\right).
\nonumber
$
}
\end{eqnarray}

\end{document}